\documentclass[sigconf]{acmart}
\usepackage{subfigure}
\usepackage{multirow}
\usepackage{xurl}

\AtBeginDocument{%
  }

\begin{document}

\title{Scaling Laws Behind Code Understanding Model}

\author{Jiayi Lin}
\authornote{Both authors contributed equally to this research.}
\email{jiayilin1024@gmail.com}
\affiliation{%
  \institution{International Digital Economy Academy}
  \streetaddress{5 Shihua Rd}
  \state{Shenzhen}
  \country{China}
  \postcode{518017}
}

\author{Hande Dong}
\authornotemark[1]
\email{donghd66@gmail.com}
\affiliation{%
  \institution{International Digital Economy Academy}
  \streetaddress{5 Shihua Rd}
  \state{Shenzhen}
  \country{China}
  \postcode{518017}
}

\author{Yutao Xie}
\email{yutaoxie@idea.edu.cn}
\affiliation{%
  \institution{International Digital Economy Academy}
  \streetaddress{5 Shihua Rd}
  \state{Shenzhen}
  \country{China}
  \postcode{518017}
}

\author{Lei Zhang}
\email{leizhang@idea.edu.cn}
\affiliation{%
  \institution{International Digital Economy Academy}
  \streetaddress{5 Shihua Rd}
  \state{Shenzhen}
  \country{China}
  \postcode{518017}
}

\renewcommand{\shortauthors}{Lin et al.}

\begin{abstract}
  The scaling law is becoming a fundamental law in many machine learning areas. That is, 
  test error falls off with the power law when increasing training data, model size, and computing resource. 
  However, whether this law is suitable for the task of code understanding is not well studied, 
  and most current language models for code understanding are about 100M parameters, which are relatively ``small'' compared to large language models. 
  In this paper, we conduct extensive experiments to investigate the scaling law for the code understanding task 
  by varying training data, model size, and computing resource. 
  We validate that the test error of code understanding models falls off with the power law when using larger models, 
  indicating that the scaling law is suitable for the code understanding task. 
  Besides, we apply different scales of models to two downstream code understanding tasks, 
  and find that the performance increases with larger scale of models. 
  Finally, we train a large-scale code understanding model named CoLSBERT with 1.5B parameters on a large dataset using more computing resource, 
  which outperforms previous work by a large margin. 
  We will release our code and the CoLSBERT model when our paper is published.
\end{abstract}


\maketitle

\section{Introduction}
\label{Introduction}

Test errors consistently exhibit a power law decline with increasing training data, model size, and computing resource~\cite{DBLP:conf/nips/SorscherGSGM22}. 
This empirical phenomenon, known as the scaling law, broadly holds in many areas, encompassing generative language models, speech processing, and translation tasks~\cite{DBLP:journals/corr/abs-2001-08361, DBLP:journals/corr/abs-2102-01293, DBLP:conf/emnlp/GordonDK21}. 
This law signifies a direct path to enhancing model performance by increasing the scale of the constituent components, including data volume, model parameters, and computing capacity~\cite{DBLP:journals/corr/abs-2010-14701}.
In recent years, we have witnessed an explosion in the development of increasingly massive models. 
Noteworthy examples include DALLE, boasting an astounding 12 billion parameters dedicated to image generation~\cite{DBLP:conf/icml/RameshPGGVRCS21}, 
Codex, endowed with an equally impressive 12 billion parameters tailored for code generation~\cite{DBLP:journals/corr/abs-2107-03374}, 
and GPT-3, wielding a staggering 175 billion parameters exclusively designed for text generation~\cite{DBLP:conf/nips/BrownMRSKDNSSAA20}. 
These colossal models have attracted substantial attention and recognition within both academic and industrial communities~\cite{DBLP:journals/corr/abs-2303-18223}.

The current state-of-the-art approach for code comprehension involves the pre-training of a transformer-encoder on a diverse array of tasks, 
including the ``mask then predict'' task, wherein the model predicts the masked tokens~\cite{DBLP:conf/emnlp/FengGTDFGS0LJZ20, DBLP:conf/iclr/GuoRLFT0ZDSFTDC21, DBLP:conf/acl/GuoLDW0022}. 
Despite its undeniable success, it is noteworthy that, to the best of our knowledge, no prior research has delved into the scaling law governing this method. 
Consequently, the question remains open as to whether straightforwardly increasing the model scale can yield further improvements.
Besides, we find that most pre-traineded models for code understanding presently employ approximately 100M parameters and are pre-traineded on the CodeSearchNet dataset, 
which comprises less than 5GB data~\cite{DBLP:conf/emnlp/FengGTDFGS0LJZ20, DBLP:conf/iclr/GuoRLFT0ZDSFTDC21, DBLP:conf/icse/NiuL0GH022, DBLP:conf/emnlp/LiGGLSQJCD22}. 
From the perspective of generic large language models, these model sizes and training data volumes might be considered relatively ``small''.
Thus, it is desirable to investigate whether the scaling law still holds for the code understanding task. 
Should this law persist, it opens up substantial potentials for improving code understanding models by increasing their scales.

To this end, we conduct extensive experiments to validate the scaling law in the code understanding models. 
Specifically, we pre-trained the transformer-encoder from scratch with the ``mask then predict'' task on different scales, 
and analyze the error of the test set to explore the law of test error in terms of the scales. 
The dimensions of scale can be delineated across three principle facets. 
1) \textbf{Training Data}: We introduce \textit{The Stack} dataset, which stands as the largest repository of code-related data presently available~\cite{DBLP:journals/corr/abs-2211-15533}, to pre-trained the code understanding model. 
\textit{The Stack} dataset significantly surpasses the scale of the previously prevalent CodeSearchNet, commonly employed for pre-training code understanding models previously.
From this extensive corpus, we sample data across varying scales and subsequently train the transformer-encoder models with these data respectively. 
2) \textbf{Model Size}. We manipulate various architectural dimensions within the transformer-encoder model, including hidden feature size, intermediate size, attention heads, and hidden layers, 
thereby influencing the total number of model parameters, i.e., the model size.
3) \textbf{Computing Resource}. We train the transformer-encoder with different iterations, 
to let the model be trained with different numbers of tokens. 
Through a comprehensive array of experiments, we have discerned that the test error exhibits a consistent decline, 
following a power-law distribution, contingent upon alterations in training data, model size, and computational resources. 
Consequently, our empirical findings affirm the validity of the scaling law within the domain of code understanding tasks.

Next, we explore whether employing a model with larger scale translates to enhanced performance in downstream code understanding tasks. 
Specifically, we aim to discern whether a model exhibiting lower test error during the pre-training phase yields superior results when fine-tuned for downstream tasks. 
To investigate this, we choose two distinct downstream tasks, namely code search and clone detection, 
and fine-tune the pre-traineded model with different scales on the two tasks. 
Experimental findings indicate that while the model's performance improvement does not precisely mirror the power law pattern observed in test error during pre-training, 
it does exhibit incremental enhancement with larger pre-training scale. 
Thus, it is evident that enlarging the pre-training scale can effectively enhance model performance in downstream code understanding tasks.

Based on the above research, a straightforward method to improving the code understanding model is 
to enlarge the scale of the pre-traineded code understanding model from the training data, model size, and training resource aspects. 
Toward this goal, we train a large scale code understanding model known as CoLSBERT, 
featuring an impressive 1.5B parameter transformer-encoder architecture. 
CoLSBERT is meticulously trained on a {304}GB dataset comprising {351}B training tokens drawn from six of the most prominently utilized programming languages as documented in \textit{The Stack} dataset. 
Subsequently, we evaluate it on downstream code understanding tasks. 
We find that CoLSBERT outperforms previous code understanding models by a large margin, 
showcasing the superiority of our CoLSBERT model.
We will release our CoLSBERT when our paper is published. 

In summary, our paper makes the following contributions: 
\begin{itemize}
    \item We validate that the scaling law holds when it comes to pre-trained the transformer-encoder for the code understanding task from the experimental perspective. 
    \item We show that the model with larger scale in the pre-training can perform better in the downstream code understanding tasks, which point out a straightforward approach to enhancing the model by enlarging the scale in the pre-training stage. 
    \item We train a 1.5B model for code understanding task on large data and training resource, validate its effectiveness, and will open-source the model to public. 
\end{itemize}

\section{Preliminary}

In this section, we begin by providing an overview of the transformer architecture, encompassing both the transformer-encoder and transformer-decoder components. 
Subsequently, we delve into the introduction of two distinct pre-training tasks, namely ``mask then predict'' and ``next token prediction''. 
The two tasks serve as the foundation respectively for training BERT models, facilitating comprehension, and GPT models, promoting generation. 
Lastly, we present the scaling law observed in neural language models, a phenomenon empirically affirmed through numerous experiments.

\subsection{Transformer Architecture} 
The Transformer architecture has emerged as the predominant model to date~\cite{DBLP:journals/csur/TayDBM23}. 
It employs a stacking mechanism of multiple transformer blocks, each comprising a self-attention layer and an MLP layer. 
Two primary self-attention structures are identified:

\begin{itemize}
\item \textbf{Full attention}~\cite{DBLP:conf/naacl/DevlinCLT19, DBLP:conf/emnlp/FengGTDFGS0LJZ20}. 
This bidirectional attention operates between every pair of tokens in the sequence. 
Notably, both subsequent and preceding tokens can assimilate information from each other. 
The robust information encoding capability afforded by full attention designates the model as a transformer-encoder, specialized for tasks involving understanding.
\item \textbf{Masking attention}~\cite{DBLP:conf/nips/BrownMRSKDNSSAA20, DBLP:journals/corr/abs-2107-03374}. 
In this directional attention model, interactions occur exclusively from left to right within the token sequence. 
Consequently, only the subsequent token can integrate information from preceding tokens, creating a unidirectional flow. 
Masking attention aligns with standard causal language modeling and is denoted as a transformer-decoder, tailored for tasks centered around generation.
\end{itemize}

\subsection{pre-training Tasks}
The primary factor contributing to the remarkable success of the Transformer model lies in the efficacy of the pre-training technique, 
which aptly trains the model using an extensive unlabeled dataset through the self-supervised learning method~\cite{DBLP:journals/csur/TayDBM23}. 
The prevalent pre-training tasks, notably the ``mask then predict'' and ``next token prediction'' tasks, are key to this success.

\begin{itemize}
    \item \textbf{mask then predict}:
Primarily employed for training the transformer-encoder, this task involves masking a portion of tokens and predicting them based on the remaining tokens in the sequence~\cite{DBLP:conf/emnlp/FengGTDFGS0LJZ20, DBLP:conf/iclr/GuoRLFT0ZDSFTDC21, DBLP:conf/acl/GuoLDW0022}. 
Specifically, a subset of tokens is randomly sampled and replaced with a special token, i.e., [MASK]. 
The objective is to predict the original tokens for the masked tokens, incorporating information from both left and right tokens in the sequence. 
This aligns with the bidirectional attention in the transformer-encoder architecture, and a transformer-encoder pre-traineded with this task is commonly referred to as BERT~\cite{DBLP:conf/naacl/DevlinCLT19}.
    \item \textbf{next token prediction}:
Tailored for training the transformer-decoder~\cite{DBLP:journals/corr/abs-2107-03374, DBLP:journals/corr/abs-2305-06161}, this task aims to predict the next token based on preceding tokens in the sequence. 
Notably, the prediction relies solely on the left tokens without considering the right tokens in the sequence, corresponding to the unidirectional attention in the transformer-decoder architecture. 
A transformer-decoder pre-traineded with this task is consistently termed GPT~\cite{DBLP:conf/nips/BrownMRSKDNSSAA20}.
\end{itemize}

In this paper, our primary focus is on the code understanding task, specifically analyzing the BERT model, 
i.e., the transformer-encoder trained with the ``mask then predic'' task on the code dataset.

\subsection{Scaling Law in Language Model}

\begin{figure}[htbp]
  \vskip 0.2in
  \begin{center}
  \centerline{\includegraphics[width=\columnwidth]{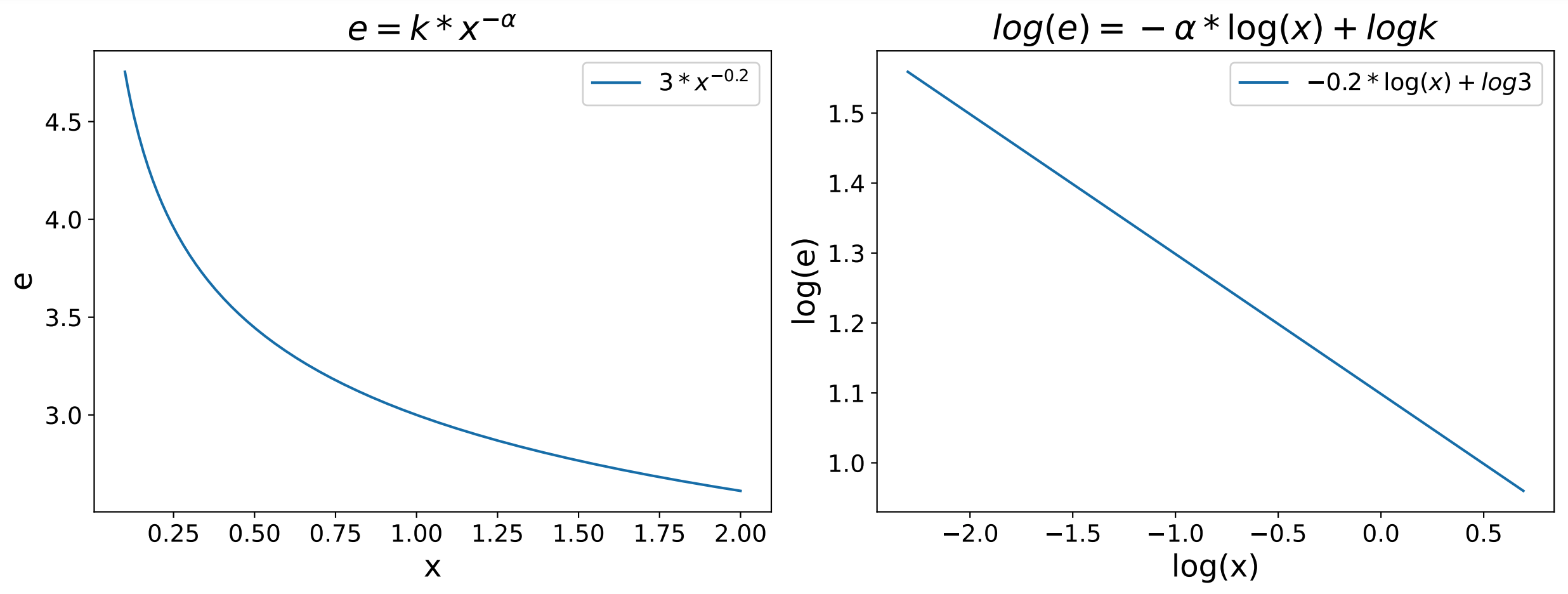}}
  \caption{An example of the power-law with the log-log plot.}
  \label{log-log}
  \end{center}
  \vskip -0.2in
\end{figure}

The scaling law has emerged as a foundational principle in the field of neural networks~\cite{DBLP:journals/corr/abs-2010-14701}, 
particularly in the domain of language models~\cite{DBLP:journals/corr/abs-2001-08361}. 
According to this law, the test error exhibits a power-law decay as either the model size, training data, and computing resources increases~\cite{DBLP:journals/corr/abs-2203-15556}. 
This relationship can be expressed through the formulation:
\begin{equation}
    e = kx^{-\alpha},
\end{equation}
where $e$ represents the test error, $x$ signifies the scale of either the model size, training data, and computing resources, $\alpha$ denotes the power-law factor, and $k$ is a scaling factor. 
Distinct values for $\alpha$ and $k$ arise when considering different elements among model size, training data, and computing resources. 
By applying logarithmic operations to the equation, we obtain $log e = -\alpha log x + log k$. 
Consequently, on a log-log plot, the test errors $e$ exhibit linearity with the training scale $x$, as illustrated in Figure~\ref{log-log}.

Although the scaling law has been validated in many different areas, inspired some researchers to train tremendous models, 
it is acutally a empirical law, 
indicating that extra experiments should be done to validate whether it holds on other task to draw a strict conclusion. 
However, no previous work explores the scaling law about code understanding task, and there lacks work which takes efforts to train large model in this task. 
Thus, we mainly focus on this research question in this paper.

\section{Scaling Law in Code Understanding Model}
\label{pre-training}

In this section, we delve into the scaling law within code understanding models. 
Initially, we present the research method and implementation details employed in this section. 
Subsequently, we present our findings concerning the scaling law in code understanding, 
specifically validating the impact of training data, model size, and computing resources, respectively.

\subsection{Method and Implementation}

We employ the transformer-encoder architecture as the code understanding model, utilizing the ``mask then predict'' task for model training. 
Varied training data, model sizes, and computing resources are manipulated to train models of different scales, and the test error is evaluated across these models. 
To explore the scaling law in three dimensions, we establish a minimum of four different scales for each dimension. 
The test error, computed using the ``mask then predict'' task, is formulated as follows:
\begin{equation}
    e = -\sum_i log(p_i | X^{mask}), 
    \label{MLM loss}
\end{equation}
where $X^{mask}$ represents the masked sequence, and $p_i$ is the probability of the masked tokens predicted by the transformer-encoder model. 
Evaluation experiments are conducted 50 times for each model, utilizing 10,000 random test samples in each iteration, and we display the mean values. 
In Appendix~\ref{uncertainty}, we demonstrate that, despite the random nature of masked tokens, the test loss consistently stabilizes under this setting.


\begin{table*}[t]
\caption{Pre-training data statistics for CodeSearchNet and The Stack. We show the number of functions. }
\label{table:data_statistics}
\begin{center}
\begin{small}
\begin{sc}
\resizebox{0.7\textwidth}{!}{
\begin{tabular}{ccccccc}
    \toprule
    Datatset      & Python & Java    & Go      & Php     & Javascript & Ruby \\
    \midrule
    CodeSearchNet & 412,176  & 454,451 & 317,832  &  523,712  & 123,889  & 48,791  \\
    \midrule
    The Stack (CSN-1x) & 412,176  & 454,451  & 317,832  & 523,712   & 123,889  & 48,791  \\
    \midrule
    The Stack (CSN-2x) & 824,352 & 908,902  & 635,664  &  1,047,424  & 247,778  & 97,582  \\
    \midrule
    The Stack (CSN-4x) & 1,660,123  & 1,829,223  & 1,282,747  & 2,106,247  & 506,795  & 138,069  \\
    \midrule
    The Stack (CSN-8x) & 3,347,859  & 3,686,059  & 2,593,701  & 4,240,151   & 1,041,563  & 138,069  \\
    \bottomrule
\end{tabular}
}
\end{sc}
\end{small}
\end{center}
\end{table*}

\textbf{Dataset} \quad
We mainly use two datasets to train and evaluate.
1) \textbf{\textit{CodeSearchNet}}~\cite{DBLP:journals/corr/abs-1909-09436}, which collects (comment, code) from Github in 2019, encompassing six common programming languages. 
Subsequent refinements to CodeSearchNet involve the removal of low-quality data based on specific criteria~\cite{DBLP:conf/iclr/GuoRLFT0ZDSFTDC21}. 
Currently, it stands as the primary dataset for training code understanding models, 
including CodeBERT~\cite{DBLP:conf/emnlp/FengGTDFGS0LJZ20}, GraphCodeBERT~\cite{DBLP:conf/iclr/GuoRLFT0ZDSFTDC21}, and UniXcoder~\cite{DBLP:conf/acl/GuoLDW0022}. 
The CodeSearchNet dataset is partitioned into training, validation, and test sets~\cite{DBLP:journals/corr/abs-1909-09436}. 
We use the CodeSearchNet-training to train the model, and use the combination of the validation and test set of the CodeSearchNet to evaluate the test loss. 
2) \textbf{\textit{The Stack}}~\cite{DBLP:journals/corr/abs-2211-15533}, which aggregates data with an open license from Github spanning the years 2015 to 2022. 
Employing various strategies such as deduplication and consideration of file line numbers, the dataset is meticulously filtered to enhance data quality~\cite{DBLP:journals/corr/abs-2301-03988, DBLP:journals/corr/abs-2305-06161}. 
In this paper, we mainly use the version of The Stack provided by StarCoder~\cite{DBLP:journals/corr/abs-2305-06161}\footnote{https://huggingface.co/datasets/bigcode/starcoderdata}. 
Despite stringent filtering, The Stack encompasses a substantial 783GB of data across 86 programming languages, surpassing the scale of CodeSearchNet. 
Following the approach outlined by Husain et al.~\cite{DBLP:journals/corr/abs-1909-09436}, we extract functions from The Stack.
The Stack is only used to train the model. 
The statistical details of these datasets are presented in Table~\ref{table:data_statistics}.

\textbf{Training Details} \quad
We train a Byte-level BPE tokenizer on the CodeSearchNet training set, following the method outlined by Liu et al.~\cite{DBLP:journals/corr/abs-1907-11692}, 
and set the vocabulary size to 50,265 as RoBERTa. 
We set the sequence length as 512, and use the AdamW optimizer to train the model. 
To facilitate training, we adopt mixed-precision training in bfloat16. 
The learning rate is set as $2e{\text{-}4}$ and decays with a linear scheduler after warming up on about 10K steps. 
We set the batch size to 256, implementing it through gradient accumulation with varying steps for different scales of models. 
We use the HGX server with 8 A100-80G GPUs to train the model. 

\begin{figure}[ht]
\begin{center}
\centerline{\includegraphics[width=0.8\columnwidth]{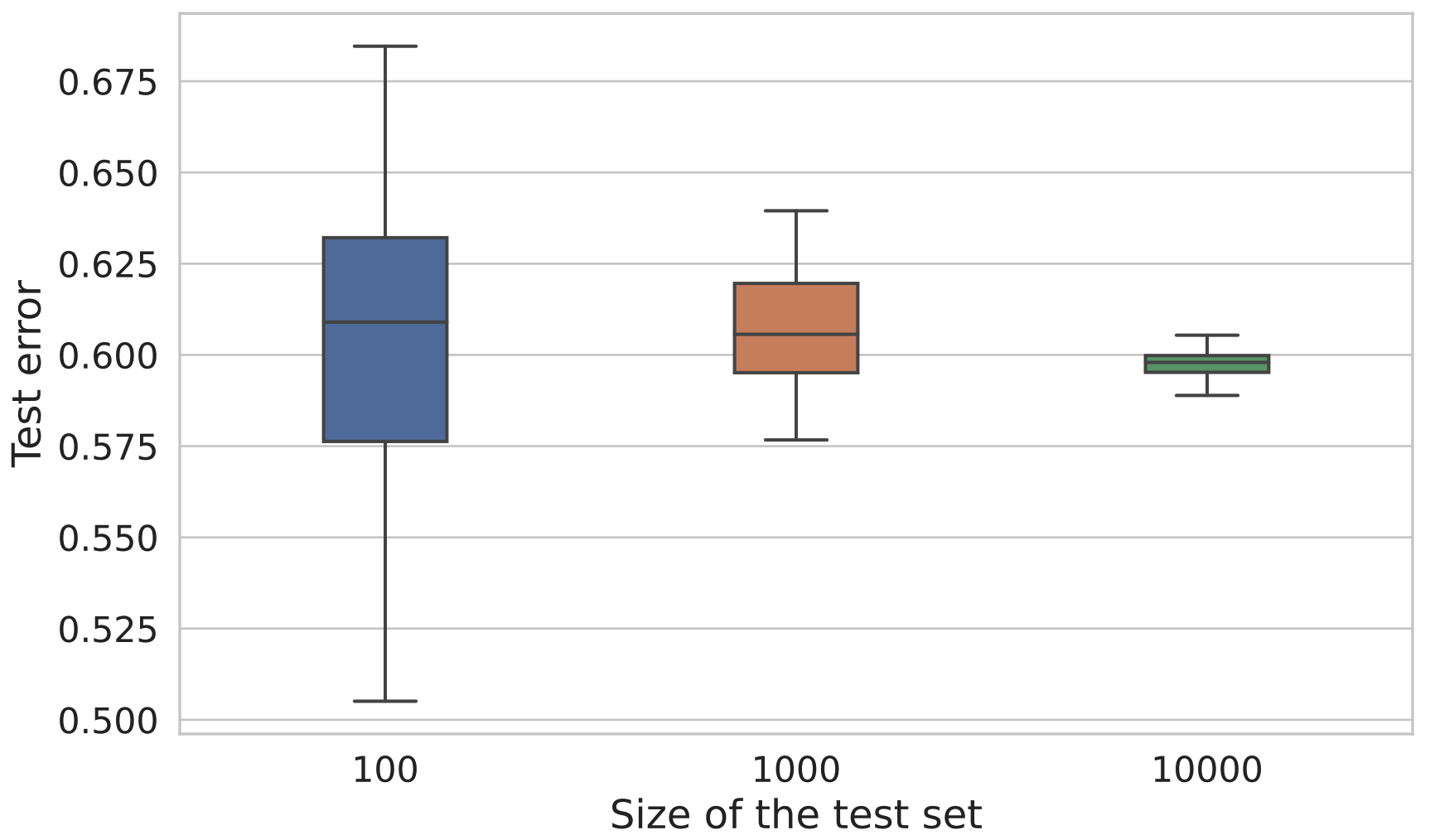}}
\caption{The test error distribution with regard to different amount of the test set.}
\label{Error variance}
\end{center}
\end{figure}

\textbf{Uncertainty of Test Error} \quad
Note that the masked tokens are randomly sampled, introducing uncertainty to the test error in Equation (\ref{MLM loss}). 
However, we can mitigate this uncertainty by expanding the size of the test set. 
Figure ~\ref{Error variance} illustrates the distribution of test errors for varying test set sizes, including 100, 1,000, and 10,000. 
To examine the impact, we conducted 50 evaluations with distinct random seeds, focusing on the ``mask then predict'' test error. 
The results are presented in a box plot, revealing that the test error stabilizes with a larger test set, with a noticeable reduction in variance when utilizing 10,000 test data. 
Notably, despite the randomness introduced by the masked tokens, the test loss remains stable. 
It is worth mentioning that the test error experiments involve a substantial dataset of 89,154 instances, surpassing the 10,000-data point, 
underscoring the stability of the test loss under these conditions.

\begin{figure*}[ht]	
	\subfigure[Training data] 
	{
		\begin{minipage}{0.32\textwidth}
			\centering          
			\includegraphics[scale=0.18]{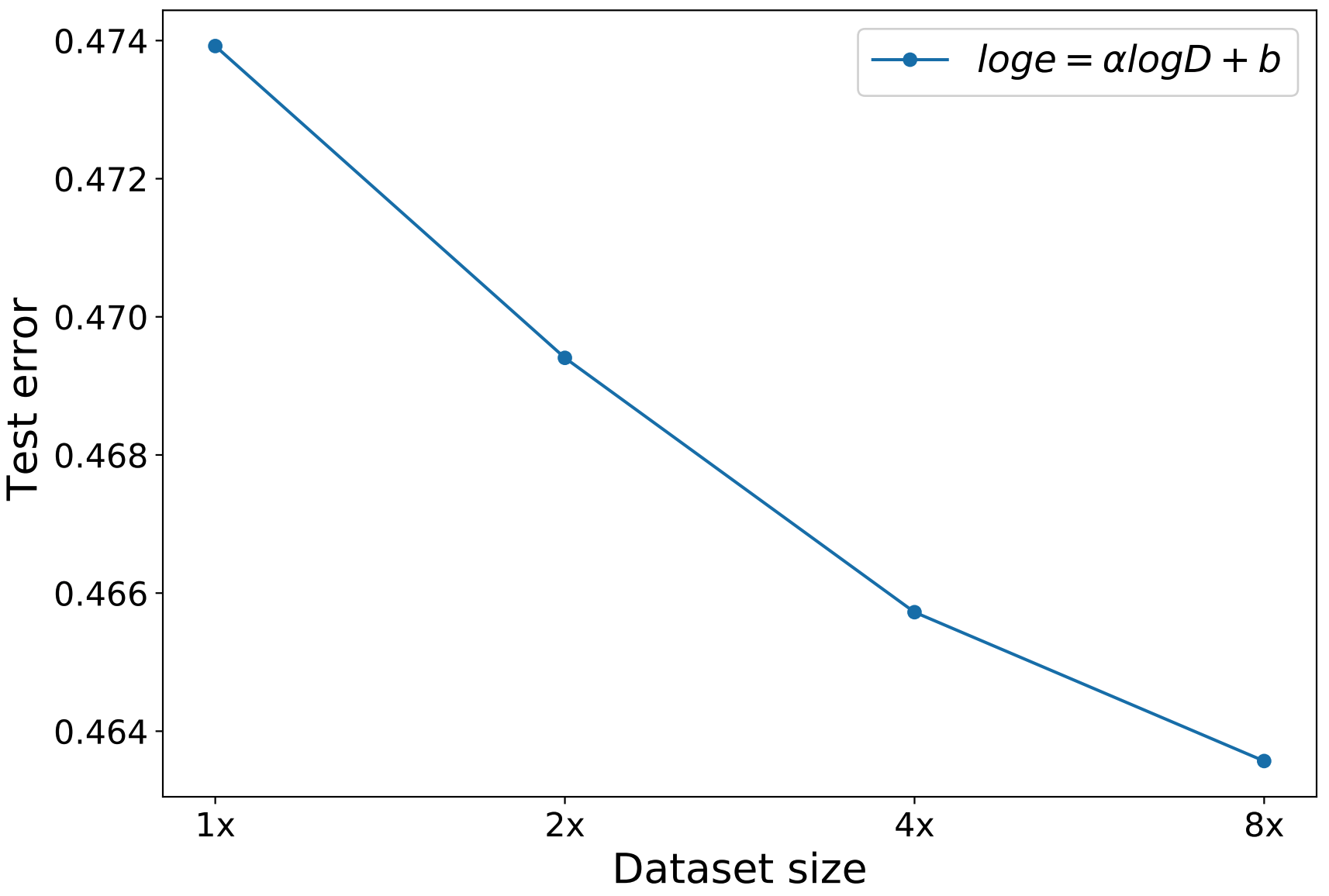}   
		\end{minipage}
	}
	\subfigure[Model size] 
	{
		\begin{minipage}{0.32\textwidth}
			\centering     
			\includegraphics[scale=0.18]{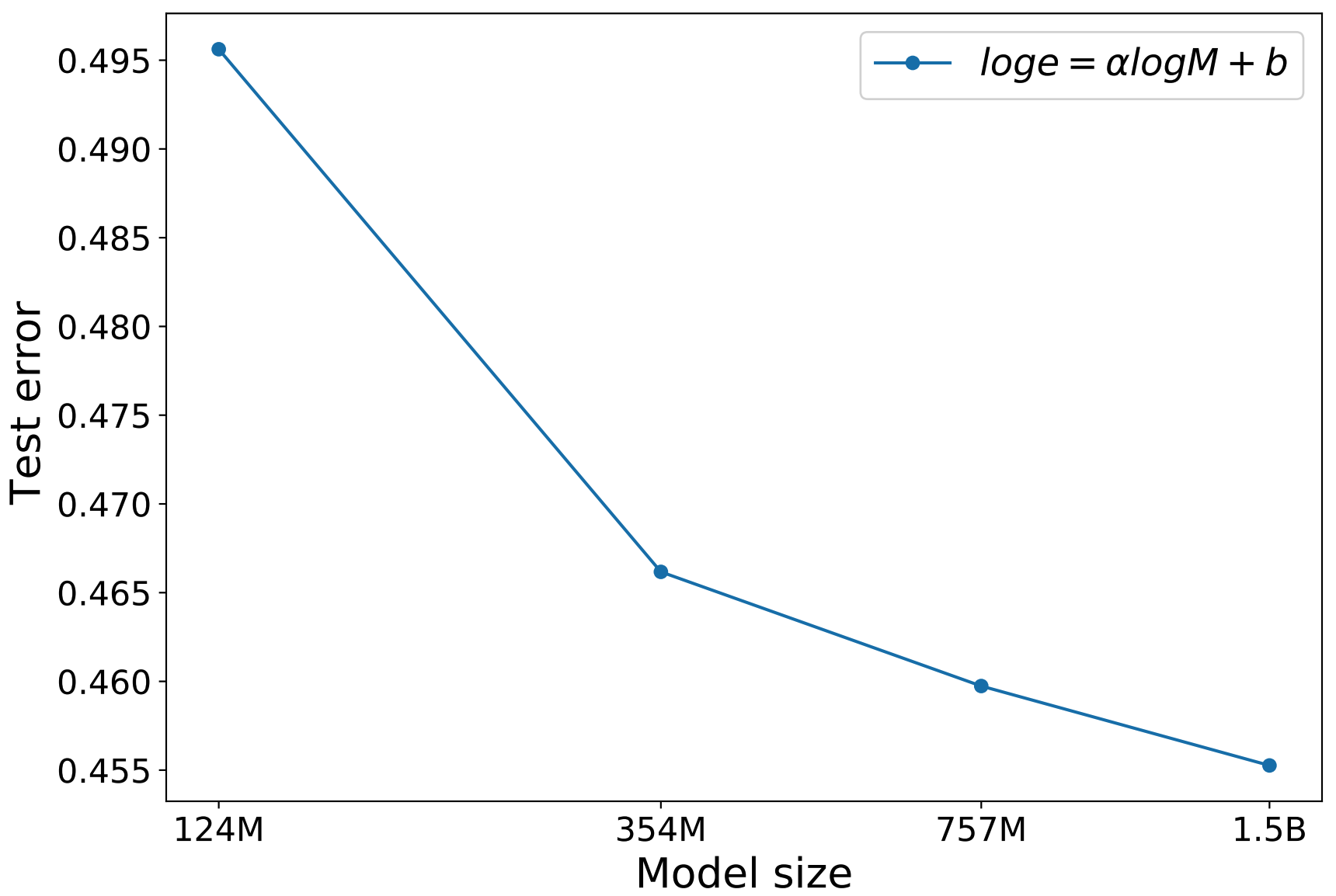}   
		\end{minipage}
	}
    \subfigure[Computing resource] 
	{
		\begin{minipage}{0.3\textwidth}
			\centering      
			\includegraphics[scale=0.18]{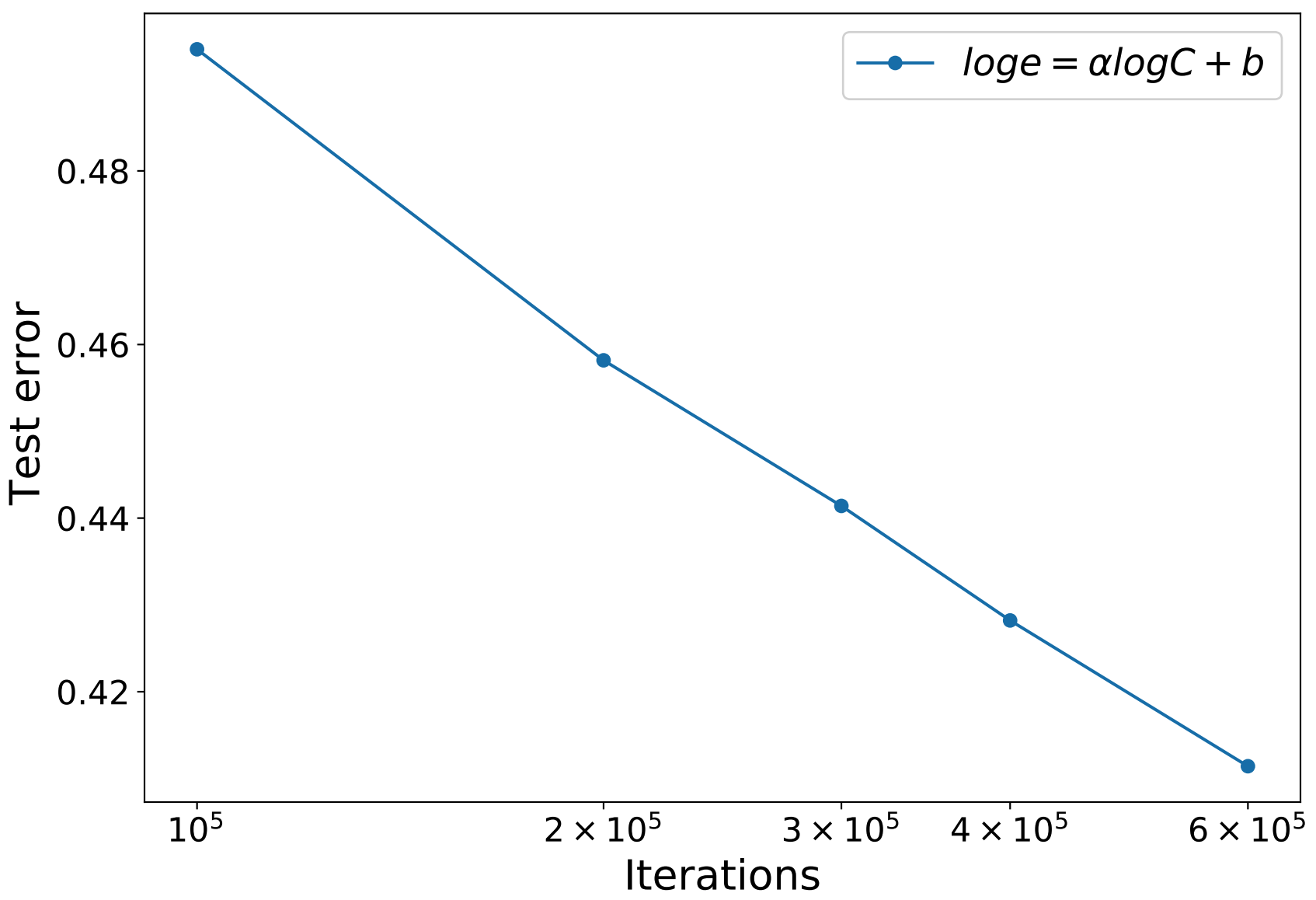}   
		\end{minipage}
	}
	\caption{The test error with regard to different scales in the code understanding task. } 
	\label{scaling law}  
\end{figure*}

\subsection{Scaling Training Data} 

We sample data from The Stack to train the model. 
As mentioned before, The Stack is much larger than CodeSearchNet-training. 
By sample different data, we can obtain different scales of combined dataset. 
In this section, we construct four scales of training data: 
CSN-1x, CSN-2x, CSN-4x and CSN-8x.
These datasets are composed of CodeSearchNet-training data scaled by factors of 1x, 2x, 4x, and 8x, respectively.
Additionally, we utilize the combination of the validation and test set of the CodeSearchNet to evaluate. 
To keep the consistency between training and evaliation, we only sample the same six programming languages as CodeSearchNet from The Stack, 
and keep the proportion of the sampled data same with the CodeSearchNet-training if the data is enough in The Stack. 
The model size is set to 124M parameters, and the training iteration is set to 100K steps, indicating that the model can see 26B tokens during training. 

The result is shown in Figure~\ref{scaling law}(a). 
From it, we observe a reduction in test error with increasing training data.
The points align well with the curve described by $log e = \alpha log D + b$, indicating a power-law relationship between test error and training data size.
Thus, the scaling law holds about the data size dimension in the code understanding task.

\subsection{Scaling Model Size} 
\label{Scaling Model Size}

\begin{table}[t]
\caption{The architectures of different scale models.}
\label{model size}
\begin{center}
\begin{small}
\begin{sc}
\resizebox{0.44\textwidth}{!}{
\begin{tabular}{ccccc}
    \toprule
    params    &    layers  &  hidden size     &   heads      &  head size    \\
    \midrule
    124M      &     12     &     768          &     12          &      64        \\
    \midrule
    354M      &     24     &     1024        &     16          &      64         \\
    \midrule
    757M      &     24     &     1536         &     16          &      96          \\
    \midrule
    1.5B        &     32     &     1920        &     20          &     96           \\
    \bottomrule
\end{tabular}
}
\end{sc}
\end{small}
\end{center}
\end{table}

We train the transformer-encoder models with different model size. 
The model size of the transformer-encoder depends on number of hidden layers, hidden size, and attention heads. 
We explore four different scales for the model size: 124M, 354M, 757M, and 1.5B, as detailed in Table~\ref{model size}. 
Intriguingly, we note an unexpected degradation in model performance beyond 757M parameters during the experiments. 
To mitigate this, we employ Pre-LN~\cite{DBLP:conf/icml/XiongYHZZXZLWL20}, involving the reordering of layer normalization and residual connections. 
We use CodeSearchNet-training to train these models with 100K iterations. 

The results are presented in Figure~\ref{scaling law}(b). 
The test error decreases as the training data increases, roughly aligning with $log e = \alpha log M + b$, indicative of a power-law relationship between test error and model size. 
Therefore, the scaling law persists regarding the model size dimension in the code understanding task.
Notably, as the model size surpasses 354M parameters, a discernible deceleration in the decline of test error becomes evident. 
It may be because when the model becomes larger, only 1.8M samples are insufficient to fully train the model. 
Using a larger training set holds promise to alleviate this issue.

\subsection{Scaling Computing Resource} 

Here, we examine test loss variations concerning different computing resources. 
Employing the architecture of the 124M parameter model in section~\ref{Scaling Model Size} and CodeSearchNet-training as the training data, 
we conduct training across varying iterations, signifying increased computing resources. 
We analyze performance across five distinct computing resources, detailed in Table~\ref{computing resource}.

\begin{table}[t]
\caption{Different scale of computing resources.}
\label{computing resource}
\begin{center}
\begin{small}
\begin{sc}
\begin{tabular}{cccc}
    \toprule
    iteration    &  seen tokens    &   flops     &   hours       \\
    \midrule
    100K         &    26B          &   $1.95e\!+\!19$          &   12.25       \\
    \midrule
    200K         &    52B          &   $3.9e\!+\!19$          &    24.6         \\
    \midrule
    300K         &    78B          &   $5.85e\!+\!19$          &    36.95         \\
    \midrule
    400K         &    104B         &   $7.8e\!+\!19$          &    49.3        \\
    \midrule
    600K         &    156B         &   $1.17e\!+\!20$          &     74        \\
    \bottomrule
\end{tabular}
\end{sc}
\end{small}
\end{center}
\end{table}

The results are presented in Figure~\ref{scaling law}(c). 
Notably, we observe a reduction in test error as computing resources increase, fitting well with the expression $log e = \alpha log C + b$. 
Therefore, the scaling law also holds about the computing resource in the code understanding task. 

\noindent \textbf{Brief summary:} \quad
Overall, our examination across three dimensions—training data, model size, and computing resources—reveals a consistent power-law relationship, 
indicating a decline in test error with increasing values in each dimension. 
Thus, the scaling law remains applicable in the realm of code understanding.

\section{Downstream Tasks Evaluation}

The pretrained transformer-encoder models have to be finetuned to be used in downstream tasks, 
indicating a gap between the test error and real applications. 
Thus, it is still unclear whether larger scale benefits to downstream tasks in code understanding task despite the scaling law discussed in the last section. 
In this section, we evaluate the code understanding model with different scales on downstream tasks. 
We first introduce our research method, and then present the experimental result about the code search and clone detecion respectively. 

\subsection{Method} 
In last section, we have trained the transformer-encoder for code understanding with different training data, model size, and computing resource. 
Here, we fine-tune these models with different scales on the two downstream tasks, namely code search and clone detection. 
We obtain the well-trained models and then evaluate their performance on downstream tasks. 
We will introduce these details, datasets, and evaluation metrics in section~\ref{model}. 
In this section, we mainly focus on the relative performance of these models.

\subsection{Code Search} 

\begin{figure*}[ht]	
	\subfigure[Training data] 
	{
		\begin{minipage}{0.32\textwidth}
			\centering          
			\includegraphics[scale=0.18]{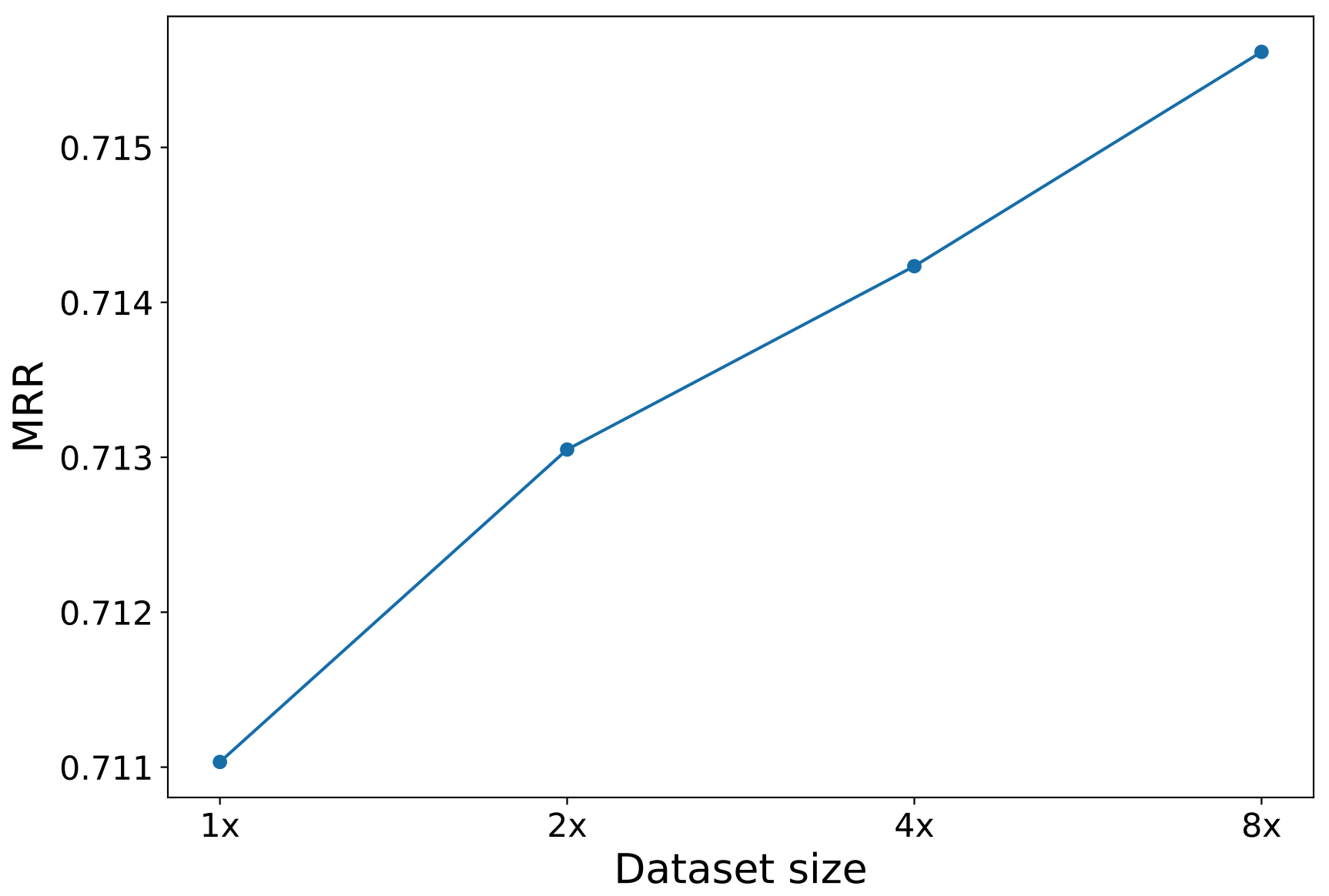}   
		\end{minipage}
	}
	\subfigure[Model size] 
	{
		\begin{minipage}{0.32\textwidth}
			\centering     
			\includegraphics[scale=0.18]{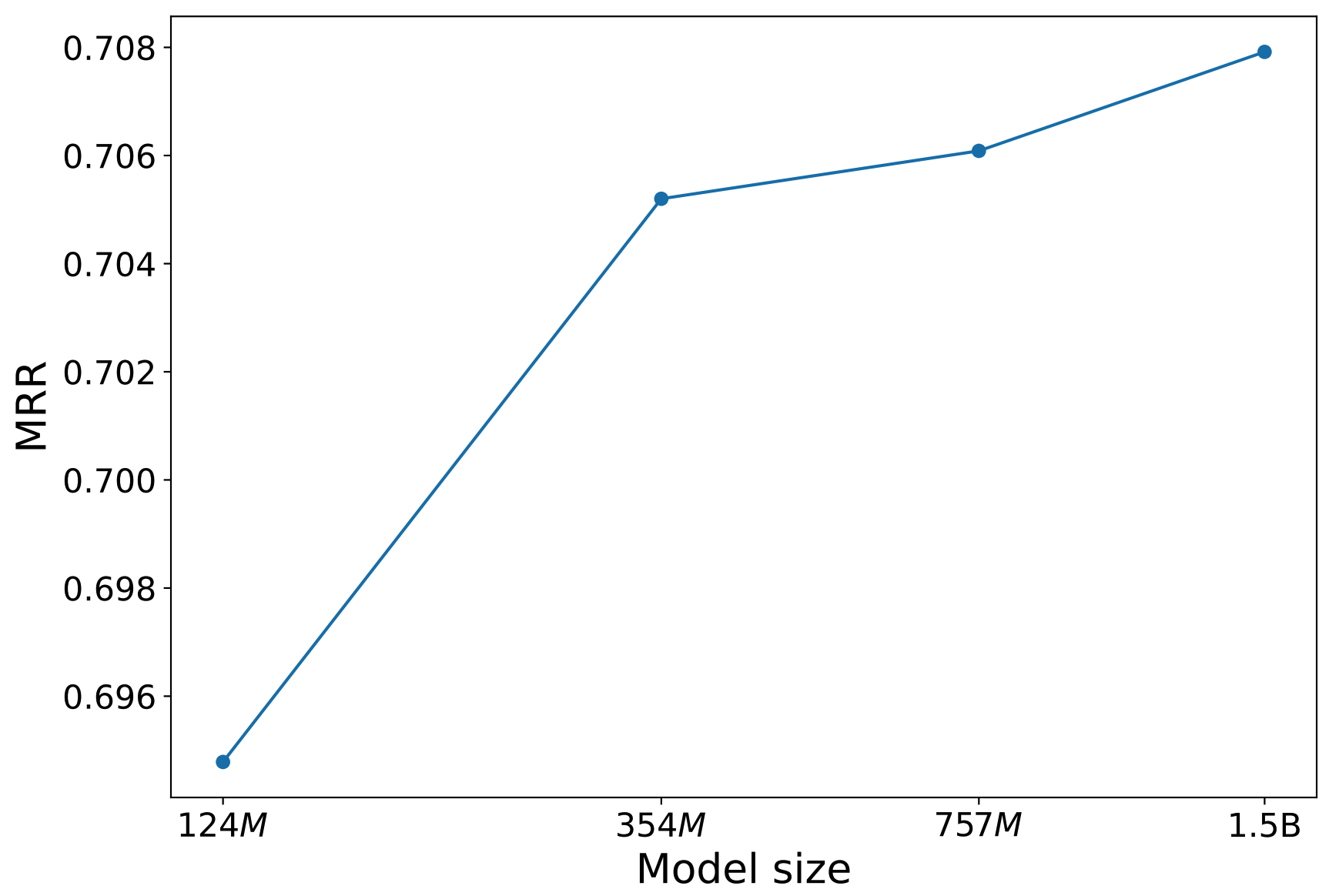}   
		\end{minipage}
	}
    \subfigure[Computing resource] 
	{
		\begin{minipage}{0.3\textwidth}
			\centering      
			\includegraphics[scale=0.18]{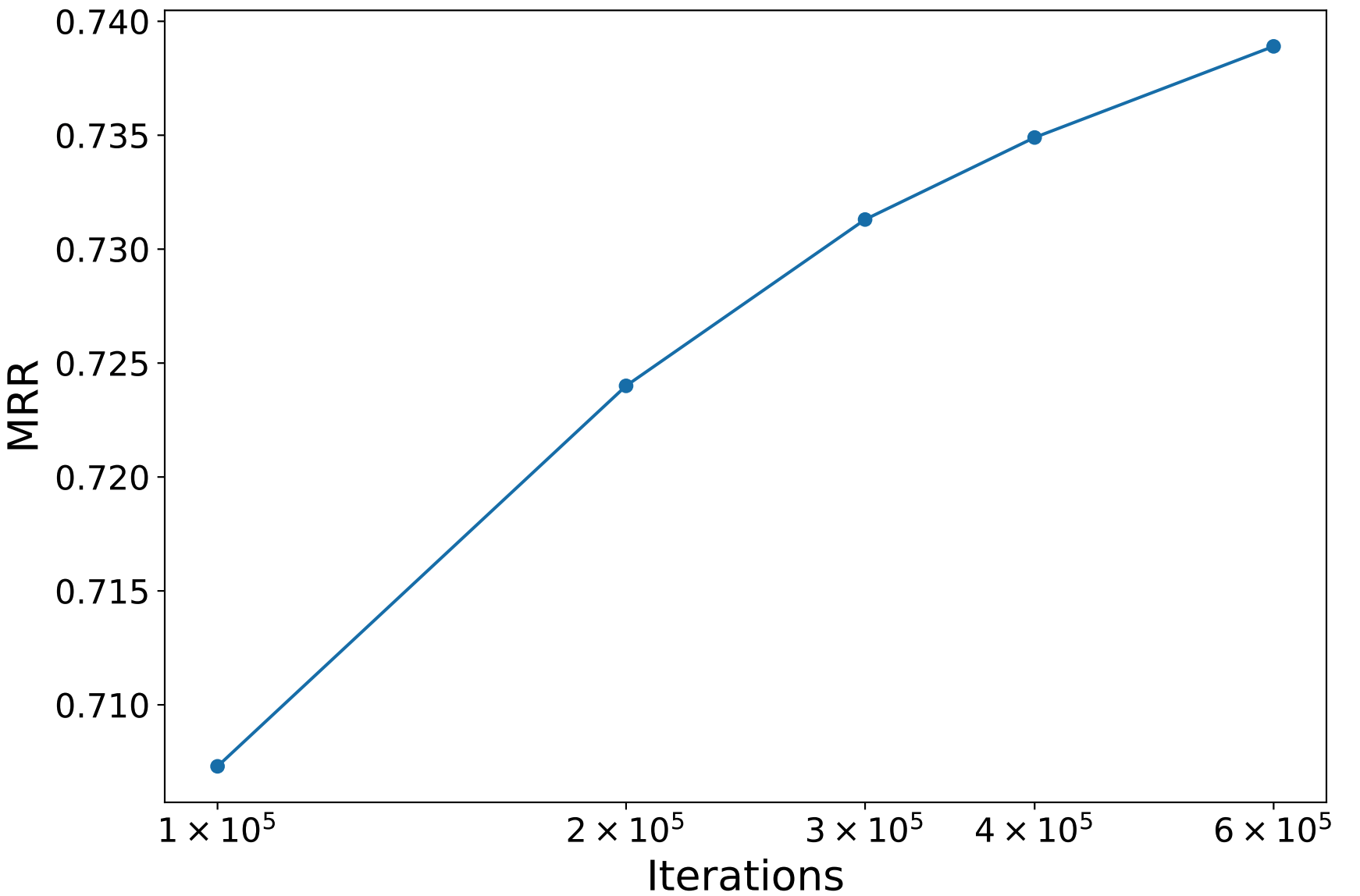}   
		\end{minipage}
	}
	\caption{Performance of different scaling models on the code search task. (a) Different pre-training data; 
    (b) Different model sizes; (c) Different pre-training computing resources. 
    The x-axis is $log(scale)$, and the y-axis is MRR on CodeSearchNet.} 
	\label{code search}  
\end{figure*}

Code search, a prevalent task in code understanding, involves seeking relevant code snippets in response to specific queries. 
we fine-tune pre-traineded models of varying scales for the code search task and evaluated their performance on the test set. 
The result is illustrated in Figure~\ref{code search}. 
Figure~\ref{code search}(a) illustrates the impact of different pre-training data, 
Figure~\ref{code search}(b) showcases the influence of varying model sizes, 
and Figure~\ref{code search}(c) delves into the effects of different pre-training computing resources. 
We can observe that for CodeSearchNet~\cite{DBLP:journals/corr/abs-1909-09436} dataset, the performance improve with either training data, model size and computing resource increasing. 
Thus, the code understanding model with larger scale performs better in the code search task.

\subsection{Clone Detection} 

\begin{figure*}[ht]	
	\subfigure[Training data] 
	{
		\begin{minipage}{0.32\textwidth}
			\centering          
			\includegraphics[scale=0.18]{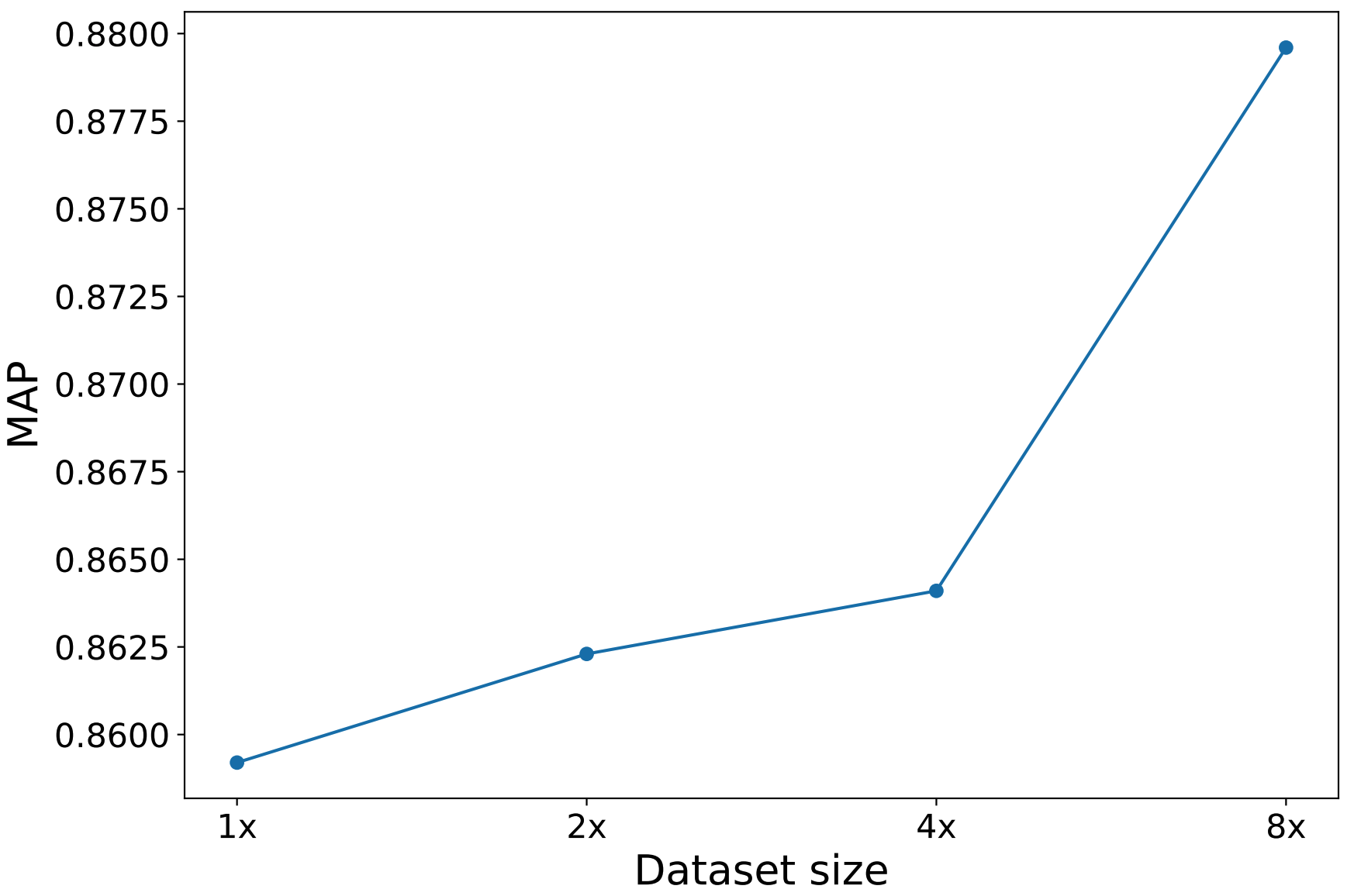}   
		\end{minipage}
	}
	\subfigure[Model size] 
	{
		\begin{minipage}{0.32\textwidth}
			\centering     
			\includegraphics[scale=0.18]{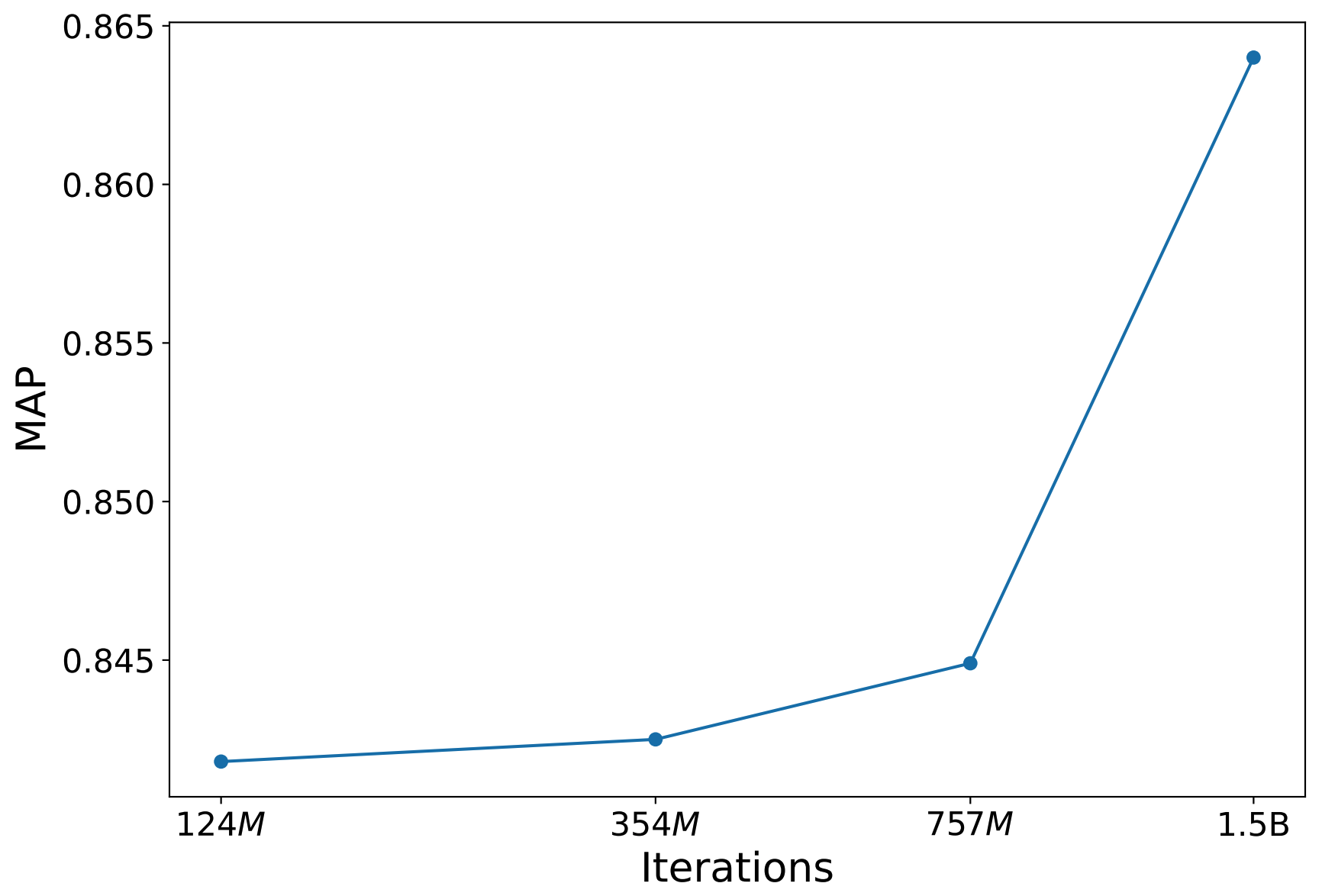}   
		\end{minipage}
	}
    \subfigure[Computing resource] 
	{
		\begin{minipage}{0.3\textwidth}
			\centering      
			\includegraphics[scale=0.18]{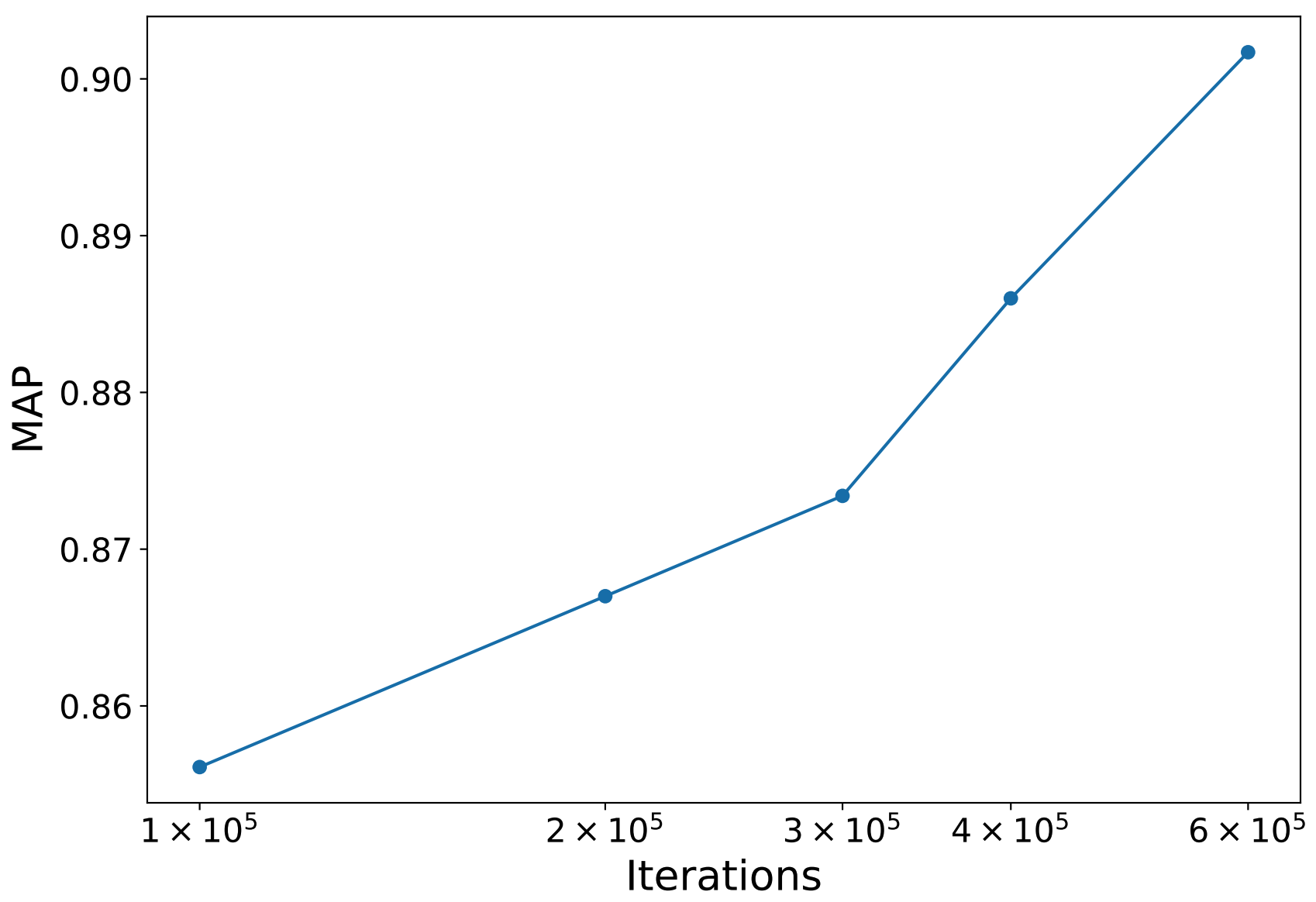}   
		\end{minipage}
	}
	\caption{Performance of different scaling models on the clone detection task. (a) Different pre-training data; 
    (b) Different model sizes; (c) Different pre-training computing resources. 
    The x-axis is $log(scale)$, and the y-axis is MAP@R for POJ-104 dataset.} 
	\label{clone detection}  
\end{figure*}

Clone detection is another common code understanding task, involves identifying similar code snippets. 
We fine-tune these pre-traineded models with different scales on clone detection task, and evaluate them on the test set. 
The results, depicted in Figure~\ref{clone detection}, offer insights into the impact of different factors—training data, model size, and computing resource. 
We can oberseve that the performance improves with either training data, model size and computing resource increasing on POJ-104~\cite{DBLP:conf/aaai/MouLZWJ16} dataset. 
Intriguingly, the observed performance enhancement exhibits a turning point, reminiscent of emergent capabilities seen in large language models. 
Thus, larger scaling code understanding model performs better in the clone detection task. 

\noindent \textbf{Brief summary:} \quad
Overall, we fine-tune pre-traineded models in section~\ref{pre-training} on two downstream tasks and datasets.
By comparing their performance, we observe superior results with larger-scaled models.
Therefore, enhancing the model scale during the pre-training stage yields improved overall model performance.

\section{Model and Result}
\label{model}
From previous section, the model with larger scale is more powerful and can achieve better performance. 
Therefore, a straightforward way to improve the code understanding model is to increase the model scale, 
including training data, model size and computing resource. 
To this end, we train a large scaling code understanding model (CoLSBERT), 
which enlarge the training data, model size and computing resource together to achieve better performance. 
In this section, we initially present the model setting of our CoLSBERT. 
Subsequently, we conduct an evaluation of CoLSBERT's performance on two downstream tasks. 
Finally, we employ probing experiments to scrutinize whether CoLSBERT effectively encodes a set of code characteristics.

\vspace{-0.2cm}
\subsection{Model Setting}
CoLSBERT is a transformer encoder model which is pre-traineded on the code data with the ``mask then predict'' task. 
We pre-train CoLSBERT from scratch on code corpus with larger scale.
Next, we introduce the pre-training dataset and model structure of CoLSBERT. 

\textbf{Pre-training Dataset} \quad
We pre-train CoLSBERT using The Stack dataset. 
Specifically, we use six programming languages in The Stack~\cite{DBLP:journals/corr/abs-2211-15533}, including Python, Javascript, Ruby, Go, Java, and PHP. 
Given the relatively short sequence length of function bodies, adherence to the CSN approach results in many samples falling below the 512-token limit. 
To improve the training efficiency, we split the source file in The Stack to 512-length sequences.
The statistics are presented in Table~\ref{pre-train_data_statistics}. 

\begin{table}[htbp]
  \caption{Pre-training data statistics of CoLSBERT.}
  \label{pre-train_data_statistics}
  \begin{center}
  \begin{small}
  \begin{sc}
  \begin{tabular}{ccccccc}
      \toprule
      language      & training  \\
      \midrule
      Python & 36,447,316    \\
      \midrule
      Java & 49,026,511    \\
      \midrule
      Go & 17,105,788    \\
      \midrule
      Php & 36,519,577    \\
      \midrule
      Javascript & 39,808,475   \\
      \midrule
      Ruby & 4,223,305    \\
      \midrule
      Total & 183,130,972  \\
      \bottomrule
  \end{tabular}
  \end{sc}
  \end{small}
  \end{center}
\end{table}

\textbf{Model Structure} \quad
The CoLSBERT is composed of 32 transformer layers, each equipped with 20 attention heads and a hidden layer feature dimension of 1920. 
To prevent model collapse, we strategically alternate the order of layer normalization and residual connections within the transformer layers. 
In total, the CoLSBERT boasts 1.5B parameters, with 1.4B dedicated to non-embedding parameters.

\textbf{Pre-training Details} \quad
We train a Byte-Pair-Encoding vocabulary with 50,265 subword units for programming languages on CodeSearchNet dataset. 
The training regimen comprises 1.34M iterations, employing a global batch size of 512. 
Thus the model is exposed to 351B tokens throughout the training process. 
We use the AdamW optimizer to train the model.
The learning rate is set as 2e-4 and decays with a linear scheduler after warming up on about 40K steps.
We conduct the pre-training on a DGX cluster equipped with 64 40G A100 GPUs. 
The entirety of the pre-training for CoLSBERT consumes about 16 days.

\subsection{Evaluation} 
In this section, we provide a brief description of compared methods, tasks, evaluation datasets and evaluation metrics. 
More details can be found in Appendix~\ref{detail_evaluation}.

\textbf{Compared Methods} \quad
We compare our method with previous pre-traineded code understanding models.
\textbf{\textit{CodeBERT}}~\cite{DBLP:conf/emnlp/FengGTDFGS0LJZ20} undergoes pre-training involving masked language modeling and replaced token prediction tasks.
\textbf{\textit{GraphCodeBERT}}~\cite{DBLP:conf/iclr/GuoRLFT0ZDSFTDC21} engages in pre-training with additional tasks including edge prediction and node alignment.
\textbf{\textit{SyncoBERT}}~\cite{DBLP:journals/corr/abs-2108-04556} leverages multi-modal contrastive learning to augment its code comprehension capabilities.
\textbf{\textit{UniXcoder}}~\cite{DBLP:conf/acl/GuoLDW0022} adopts a diverse pre-training strategy, encompassing tasks such as cross-modal matching and language modeling.

\textbf{Tasks} \quad
We employ two downstream tasks to assess the efficacy of our model, specifically \textbf{\textit{Code Search}} and \textbf{\textit{Clone Detection}}. 
Code search endeavors to identify the most relevant code from an extensive pool of candidates based on a natural language query. 
We conduct experiments on the CodeSearchNet (CSN)~\cite{DBLP:journals/corr/abs-1909-09436} dataset, encompassing six programming languages. 
The Mean Reciprocal Rank (MRR) serves as the evaluation metric for this task.
Clone detection aims to identify semantically similar code segments. 
Our experimentation focuses on the POJ-104~\cite{DBLP:conf/aaai/MouLZWJ16} dataset, which is designed for retrieving semantically analogous codes when provided with a code query, 
utilizing Mean Average Precision (MAP) as the evaluation metric.

Moreover, to assess the proficiency of pre-trained model in comprehending code across surface, syntactic, structural, and semantic dimensions, 
we employ four probing tasks for model examination. 
These tasks include \textbf{\textit{Code Length Prediction (LEN)}}, \textbf{\textit{AST Node Tagging (AST)}}, \textbf{\textit{Cyclomatic Complexity (CPX)}}, and \textbf{\textit{Invalid Type Detection (TYP)}}.
The objective of LEN is to predict the length of a code sequence, while AST aims to predict the types of Abstract Syntax Tree nodes. 
CPX is designed to forecast the cyclomatic complexity of source code, and TYP focuses on distinguishing between valid and invalid code snippets. 
For these probing tasks, we utilize datasets sourced from Karmakar and Robbes~\cite{DBLP:conf/kbse/KarmakarR21}, derived from a subset of the 50K-C dataset comprising compilable Java projects.
All the probing tasks employed are classification tasks, and we measure performance using classification accuracy as the metric. 
Further details for each classification task are provided in Appendix~\ref{detail_dataset}. 

\textbf{Implementation Details} \quad
The experimental details for code search and clone detection are followed from UniXcoder~\cite{DBLP:conf/acl/GuoLDW0022}, and are the same for different scales without tuning hyper-parameters 
to avoid disturbance of these hyper-parameters in the fine-tuning stage. 
We load the pre-trained CodeBERT, GraphCodeBERT and UniXcoder from Huggingface~\footnote{https://huggingface.co/models},
reproduce the results of these models on downstream tasks, and present them in the paper. 
Notably, as the pre-trained model SyncoBERT is not publicly available, we refer directly to the reported results in the original paper~\cite{DBLP:journals/corr/abs-2108-04556}.
The experimental details for probing tasks are followed from Karmakar and Robbes~\cite{DBLP:conf/kbse/KarmakarR21}. 
Specifically, we fine-tune CodeBERT, GraphCodeBERT, UniXcoder, and CoLSBERT for code search and clone detection tasks. 
Subsequently, we extract the feature vector from the last hidden layer of the fine-tuned models to train a simple linear classifier, assessing the model's understanding of code-related information.

\subsection{RQ1: How effective is our proposed CoLSBERT?}

\begin{table*}[t]
\caption{Performance comparison of code search and clone detection.}
\label{performance}
\begin{center}
\begin{small}
\begin{sc}
\resizebox{0.95\textwidth}{!}{
\begin{tabular}{ccccccccc}
    \toprule
    \multirow{2}{*}{Model} &  \multicolumn{7}{c}{Code Search} & \multirow{2}{*}{Clone Detection} \\
    \cmidrule(lr){2-8}
                                                          & Ruby     & Javascript & Go      &  Python   & Java    & Php        & Avg.    &  \\
    \midrule
    CodeBERT~\cite{DBLP:conf/emnlp/FengGTDFGS0LJZ20}      &   68.0   &  62.6      &  89.2   &  68.5     &  68.6   &   64.2     & 70.2   &  83.79 \\
    \midrule
    GraphCodeBERT~\cite{DBLP:conf/iclr/GuoRLFT0ZDSFTDC21} &  70.6    &  65.6      &  90.0   &  70.6     &  70.0   &  65.7     &  72.1   &  85.50 \\
    \midrule
    SyncoBERT~\cite{DBLP:journals/corr/abs-2108-04556}    &  72.2    &  67.7      &  91.3    &  72.4     &  72.3    &   67.8      &  74.0    &   88.24 \\
    \midrule
    UniXcoder~\cite{DBLP:conf/acl/GuoLDW0022}             &  73.9    &  68.6      &  91.6    &  72.3     &  72.7    &   67.3       &  74.4     &   89.56 \\
    \midrule
    CoLSBERT                                              &   \textbf{76.8}  &   \textbf{72.2}  &  \textbf{92.2}    &   \textbf{75.9}  &   \textbf{75.1}  &    \textbf{70.0}    &  \textbf{77.0}   &  \textbf{92.91} \\
    \bottomrule
\end{tabular}
}
\end{sc}
\end{small}
\end{center}
\end{table*}

The experimental results are shown in Table~\ref{performance}.
Owing to the integration of data flow information in the source code, GraphCodeBERT outperforms CodeBERT. 
UniXcoder and SyncoBERT, leveraging the structural information of the source code, 
exhibit superior efficacy compared to alternative methods. 
CoLSBERT is pre-trained from scratch exclusively on The Stack dataset, employing the “mask then predict'' task only. 
CoLSBERT enlarge the training data, model size and computing resource together during pre-training.
Consequently, CoLSBERT excels among all methods and achieves state-of-the-art performance in both code search and clone detection tasks.

\subsection{RQ2: Why does our proposed CoLSBERT work?}
While CoLSBERT exhibits exceptional performance in the aforementioned scenarios, 
it falls short of explicitly elucidating the underlying reasons for its superior performance. 
Following Karmakar and Robbes~\cite{DBLP:conf/kbse/KarmakarR21},
we employ four probing tasks to examine whether CoLSBERT effectively captures crucial features relevant to code analysis. 
Due to limitations in space, we exclusively present the probe results of the code search fine-tuned model here, 
with the probe results of the clone detection fine-tuned model reserved for inclusion in Appendix~\ref{probing_code_clone}. 

The experimental results, outlined in Table~\ref{Probing}, 
unveil the outstanding performance of CoLSBERT in Code Length Prediction, achieving an accuracy of 61.21\%, significantly surpassing other models. 
This performance can be attributed to CoLSBERT's exposure to an extensive code corpus through the ``mask then predict'' task, 
enabling robust predictions of token numbers in input code sequences. 
It is noteworthy that additional pre-training tasks for other models may hurt performance in this task.

For AST Node Tagging, both CoLSBERT and CodeBERT demonstrate exceptional performance, achieving classification accuracies exceeding 85\%. 
Intriguingly, despite UniXcoder taking the Abstract Syntax Tree as partial input, its performance on this task is suboptimal. 
We explore this further in Appendix~\ref{CS_Probing_Each_Layer}.

UniXcoder excels in Cyclomatic Complexity prediction, achieving an accuracy of 34.73\%. 
The consideration of structural information during the pre-training stage contributes to this result. 
However, UniXcoder's advantage over other models in this task is not conspicuous, with a mere 0.29\% increase in accuracy compared to CoLSBERT. 
It is plausible that existing pre-trained code models have not effectively encoded the structural information of the code. 

For Invalid Type Detection, all fine-tuned models exhibit exemplary performance, with prediction accuracy rates surpassing 85\%. 
CoLSBERT particularly stands out with an impressive accuracy rate of 95.42\%, 
underscoring its proficiency in encoding semantic information within the code. 

Overall, CoLSBERT demonstrates superior performance across three probing tasks: Code Length Prediction, AST Node Tagging, and Invalid Type Detection. 
It is evident that CoLSBERT effectively encodes surface, syntactic, and semantic information of the code within the hidden layers of the model. 
The Cyclomatic Complexity prediction task stands out as the most challenging, warranting further in-depth research.

\begin{table}[t]
\caption{Probing task accuracy on code search fine-tuned models.}
\label{Probing}
\begin{center}
\begin{small}
\begin{sc}
\resizebox{0.47\textwidth}{!}{
\begin{tabular}{ccccc}
\toprule
\multirow{2}{*}{Model}                                &  LEN        &    AST       &     CPX         &  TYP       \\
\cmidrule(lr){2-5}
                                                        &  surface    &    syntactic &    structural   & semantic   \\
\midrule
CodeBERT                                                &    39.77    &   87.08      &    29.03      &    85.66       \\
\midrule
GraphCodeBERT                                           &   43.43     &    53.5       &     31.67       &    87.31       \\
\midrule
UniXcoder                                               &   49.92     &   10.21      &     \textbf{34.73}       &    88.87         \\
\midrule
CoLSBERT                                              &   \textbf{61.21}     &    \textbf{88.12}     &     34.44        &    \textbf{95.42}          \\
\bottomrule
\end{tabular}
}
\end{sc}
\end{small}
\end{center}
\end{table}

\section{Related Work}

Early work~\cite{DBLP:conf/icse/GuZ018, DBLP:conf/www/YaoPS19} that introduces deep learning to code understanding is to treat the code as natural language sequence, 
and then use language encoder (such as LSTM~\cite{DBLP:conf/icse/GuZ018}) to encode code. 
Later, some work argues that the structure of code (such as abstract syntax tree, AST~\cite{DBLP:conf/acl/HaldarWXH20}) is quite important, 
but language encoder does not explicitly model these structure~\cite{DBLP:journals/tkdd/LingWWPMXLWJ21}. 
Thus, these work proposes that the code should be represented to graph structure firstly, 
and then encoded with a graph relevant encoder~\cite{DBLP:journals/tosem/ZengYLXWGBDL23, DBLP:conf/wcre/MaYLJMXDL23}. 
Inspired by the remarkable success of pre-training in natural language processing~\cite{DBLP:conf/naacl/DevlinCLT19, DBLP:conf/acl/LewisLGGMLSZ20, DBLP:journals/jmlr/RaffelSRLNMZLL20}, 
some work introduces pre-training to the code understanding task, 
and improve the code understanding ability by a large margin~\cite{DBLP:conf/emnlp/FengGTDFGS0LJZ20, DBLP:conf/iclr/GuoRLFT0ZDSFTDC21, DBLP:conf/emnlp/0034WJH21, DBLP:journals/corr/abs-2108-04556, DBLP:conf/acl/GuoLDW0022, DBLP:conf/naacl/WangWWWZLWL22}. 
Specifically, CodeBERT~\cite{DBLP:conf/emnlp/FengGTDFGS0LJZ20} firstly introduces pre-training to code understanding, 
and pre-trained the model with masked language modeling and replace token prediction tasks; 
GraphCodeBERT~\cite{DBLP:conf/iclr/GuoRLFT0ZDSFTDC21}, UniXcoder~\cite{DBLP:conf/acl/GuoLDW0022}, and SynCoBERT~\cite{DBLP:journals/corr/abs-2108-04556} introduces some extra tasks to pre-train the model. 
Despite achieving state-of-the-art performance, 
these models are with relatively small scale, with about 100M parameters pre-traineded on the CodeSearchNet dataset. 

The superiority of large-scale models over their smaller counterparts has been empirically substantiated, 
particularly evident in the success of large language models~\cite{DBLP:journals/corr/abs-2203-15556}.
In the natural language field, OpenAI has successively released a series big models, 
from GPT-1~\cite{radford2018improving} with 117M parameters, to GPT-2~\cite{radford2019language} with 1.5B parameters, to GPT-3~\cite{DBLP:conf/nips/BrownMRSKDNSSAA20} with a remarkable 175B parameters. 
OpenAI has further extended its influence by releasing the APIs for ChatGPT~\cite{DBLP:conf/nips/Ouyang0JAWMZASR22} and GPT-4~\cite{DBLP:journals/corr/abs-2303-08774}, 
dedicated to supporting conversation, capturing widespread attention and marking a pivotal milestone in language modeling.
In the code field, 
there is a discernible trend towards increasing the size of generative models with Transformer decoder architecture. 
Notably, several models with magnitudes in the order of 10B parameters have been developed, 
including Codex~\cite{DBLP:journals/corr/abs-2107-03374} with 12B parameters, CodeGen~\cite{DBLP:conf/iclr/NijkampPHTWZSX23} with 16B parameters, 
CodeGeeX~\cite{DBLP:journals/corr/abs-2303-17568} with 13B parameters, and StarCoder~\cite{DBLP:journals/corr/abs-2305-06161} with 15.5B parameters. 
Recent advancements have pushed the boundaries even further, with models like Code Llama~\cite{roziere2023code} and DeepSeek-Coder~\cite{guo2024deepseek} reaching an impressive scale of 30B parameters. 
However, no work attmpts to enlarge the scale of code understanding model to our best knowledge, 
which motivates us to conduct this research.

\section{Conclusion and Future Work}
In this paper, we have conducted comprehensive experiments to investigate the scaling law in the code understanding task and  
provide empirical evidence to confirm the validity of the scaling law in this context. 
Furthermore, we showed that larger-scale models outperform smaller ones when being evaluated on downstream tasks. 
Based on these findings, we trained the CoLSBERT model by enlarging the model scale dedicated to code understanding. 
We subsequently validated its efficacy on tasks such as code search and clone detection. 
 
We propose three directions for future work. 
1) Although our trained CoLSBERT is much larger than previous code understanding models, it is still relatively small compared 
to generic large language models such as LLaMa~\cite{DBLP:journals/corr/abs-2302-13971} and GPT-3~\cite{DBLP:conf/nips/BrownMRSKDNSSAA20} which has shown a remarkable ability in various fields. 
Thus, training larger code understanding model is promising to further improve the performance. 
2) In this paper, we investigate the three dimensions of scale separately, including training data, model size, and computing resource. 
However, their collective impact on test error is not explored. 
Examining their interplay is crucial for balancing the three dimensions and optimizing the utilization of computing budget efficiently.
3) In fact, the power law of the scaling law is extremely weak, examplified by that the error drop from 3\% to 2\% requires an order of magnitude 
of training data, model size, and computing resource~\cite{DBLP:conf/nips/SorscherGSGM22}. 
Therefore, it is worthwhile to explore how to train the code understanding model in order to break the scaling law and make the test error drop more rapidly as the scale increases.

\bibliographystyle{ACM-Reference-Format}
\bibliography{sample-base}

\appendix

\section{Additional Experimental details}
\label{detail_evaluation}

\subsection{Detailed Compared Method Descriptions}
\label{detail_method}
\begin{itemize}
    \item \textbf{\textit{CodeBERT}}~\cite{DBLP:conf/emnlp/FengGTDFGS0LJZ20} is the first large-scale natural language-programming language
    pre-training model for code understanding. It is pre-trained with two objectives, namely Mask Language Modeling (MLM) and Replaced Token Detection (RTD).
    \item \textbf{\textit{GraphCodeBERT}}~\cite{DBLP:conf/iclr/GuoRLFT0ZDSFTDC21} is an upgraded version of CodeBERT. 
    It adds two new objectives to explore code structure information based on the pre-training objectives of CodeBERT, 
    namely Edge Prediction (EP) and Node Alignment (NA).
    \item \textbf{\textit{SyncoBERT}}~\cite{DBLP:journals/corr/abs-2108-04556} constructs positive samples from multiple views of code, 
    and subsequently leverages multi-modal contrastive learning to enhance the understanding of code.
    \item \textbf{\textit{UniXcoder}}~\cite{DBLP:conf/acl/GuoLDW0022} is a unified cross-modal pre-trained model for programming languages. 
    It takes information from two modalities, namely simplified AST and code comments, as input, and is pre-trained using MLM, unidirectional language modeling, denoising autoencoder, and two contrastive learning-related tasks.
\end{itemize}

\subsection{Detailed Task Descriptions}
\begin{itemize}
    \item \textbf{\textit{Code search}}: This task aims to find the most relevant code from a large collection of candidates given a natural language query. 
    The embeddings for both the query and code are derived by normalizing the averages of all embeddings through the Transformer encoder corresponding to their respective tokens. 
    Subsequently, a dot product is applied to the query embedding and code embedding to assess their relevance accurately.
    \item \textbf{\textit{Clone detection}}: The primary objective of this task is to identify codes with semantic similarities. 
    The computation of similarity between two codes follows the same methodology employed for gauging the similarity between a query and code in the context of code search.
    \item \textbf{\textit{Code Length Prediction (LEN)}}: This task is dedicated to predicting the length of a code sequence, a crucial attribute that encapsulates essential information. 
    Given that code sequences inherently differ in information content according to their length, the objective here is to partition code sequence lengths into five intervals, 
    thereby transforming the task into a multi-classification task focused on predicting the length interval to which a given code sequence belongs. 
    Through this exploration, our aim is to evaluate whether these models encode and capture this fundamental surface information of the code.
    \item \textbf{\textit{AST Node Tagging (AST)}}: The primary objective of this task is to forecast the categories of nodes in the Abstract Syntax Tree, 
    serving as a conceptual representation of the underlying code structure. 
    The Abstract Syntax Tree captures the hierarchical relationships embedded in the syntax structure, presenting it as a tree where each node corresponds to a structural unit in the source code. 
    During the pre-training phase, the objective of the model is to acquire a deep understanding of the grammatical structure of the code token sequence, which is a prerequisite for excelling in subsequent code comprehension tasks. 
    Recognition of Abstract Syntax Tree node tags is therefore deemed an implicit requirement for addressing the assigned code task, 
    encompassing a comprehensive set of 20 node types and framing it as a multi-classification. 
    This task serves as an evaluation of the pre-trained model's proficiency in encoding the inherent syntactic information within the code.
    \item \textbf{\textit{Cyclomatic Complexity (CPX)}}: The objective of this task is to anticipate the cyclomatic complexity of source code, an inherent feature reflecting its structural intricacies. 
    Complexity, an intrinsic characteristic of source code, arises from the count of linearly independent paths within a code segment. 
    Forecasting this complexity based solely on the token sequence presents a distinctive challenge. 
    The code's complexity is classified on a scale from 0 to 9, thereby transforming the task into a multi-classification. 
    Through this initiative, we aim to scrutinize the degree to which the structural information of the code is encoded within the hidden layers of the pre-trained model.
    \item \textbf{\textit{Invalid Type Detection (TYP)}}: The primary objective of this task is to identify the valid and invalid code snippets, 
    with the latter intentionally generated by misspelling primitive data types within the code snippet. 
    This simplification gives rise to a binary classification task, with two distinct classes - valid and invalid. 
    Through this research initiative, we aim to evaluate the ability of the model in grasping the semantics of the code, 
    even when challenged by deliberately introduced invalid semantic information within the given context.
\end{itemize}

\subsection{Detailed Evaluation Dataset Descriptions}
\label{detail_dataset}
\begin{itemize}
    \item \textbf{\textit{Code search}}: We conduct code search experiments on the CodeSearchNet (CSN)~\cite{DBLP:journals/corr/abs-1909-09436} dataset. 
    This is a large-scale benchmark dataset produced for code search task. 
    The dataset encompasses a diverse range of programming languages, including Python, Java, Go, PHP, JavaScript, and Ruby. 
    It has been widely used in previous studies~\cite{DBLP:conf/emnlp/FengGTDFGS0LJZ20, DBLP:conf/iclr/GuoRLFT0ZDSFTDC21, DBLP:journals/corr/abs-2108-04556, DBLP:conf/acl/GuoLDW0022}. 
    \item \textbf{\textit{Clone detection}}: We conduct code clone detection experiments on the POJ-104~\cite{DBLP:conf/aaai/MouLZWJ16} dataset.
    This dataset is utilized to retrieve semantically similar codes when given a code as the query. 
    The dataset originates from a pedagogical programming open judge (OJ) system, 
    where students submit their source code as a solution to a specific problem. 
    \item \textbf{\textit{Probing Tasks}}: We adopt datasets from Karmakar and Robbes~\cite{DBLP:conf/kbse/KarmakarR21} for the above four probing tasks, 
    derived from a subset of the 50K-C dataset comprising compilable Java projects.
    For the LEN task, code sequence lengths are categorized into five intervals (0-50, 50-100, etc.). 
    In the AST task, a diverse range of AST node tags is collected, divided into 20 types. 
    The cyclomatic complexity labels for the CPX task are obtained using the metrix++ tool, ranging from 0 to 9. 
    As for the TYP task, code snippet types are classified into two categories: valid and invalid.
    Each dataset for the respective tasks comprises 10,000 samples, meticulously balanced in terms of class distribution.
\end{itemize}

\subsection{Detailed Evaluation Metric Descriptions}
\begin{itemize}
    \item \textbf{\textit{Code search}}: We employ the \textit{Mean Reciprocal Rank (MRR)} as evaluation metric for code search task, a widely acknowledged measure in prior research endeavors~\cite{DBLP:conf/emnlp/FengGTDFGS0LJZ20, DBLP:conf/iclr/GuoRLFT0ZDSFTDC21, DBLP:journals/corr/abs-2108-04556, DBLP:conf/acl/GuoLDW0022}.
    MRR represents the average reciprocal rank of the correct code snippet given a query.
    \item \textbf{\textit{Clone detection}}: We use \textit{Mean Average Precision (MAP)} evaluation metric for clone detection task.
    MAP signifies the average reciprocal rank of all search results.
    \item \textbf{\textit{Probing Tasks}}:
    All probing tasks we employed are classification tasks, and we utilize classification accuracy as the metric for these tasks.
\end{itemize}

\section{Additional Results}

\subsection{Probing Results for Each Layer of the Code Search Models}
\label{CS_Probing_Each_Layer}

\begin{figure*}[ht]	
    \vskip 0.2in
	{
		\begin{minipage}{0.48\textwidth}
			\centering          
			\includegraphics[scale=0.28]{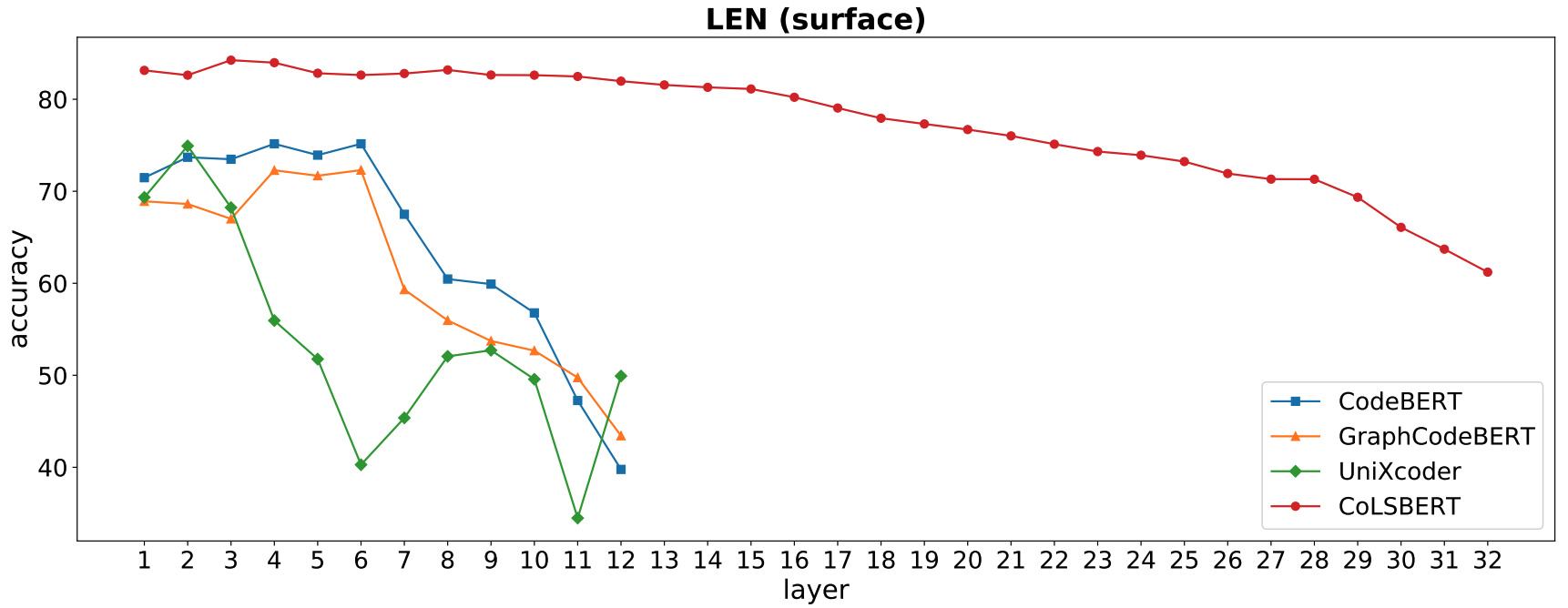}   
		\end{minipage}
	}
	{
		\begin{minipage}{0.5\textwidth}
			\centering     
			\includegraphics[scale=0.28]{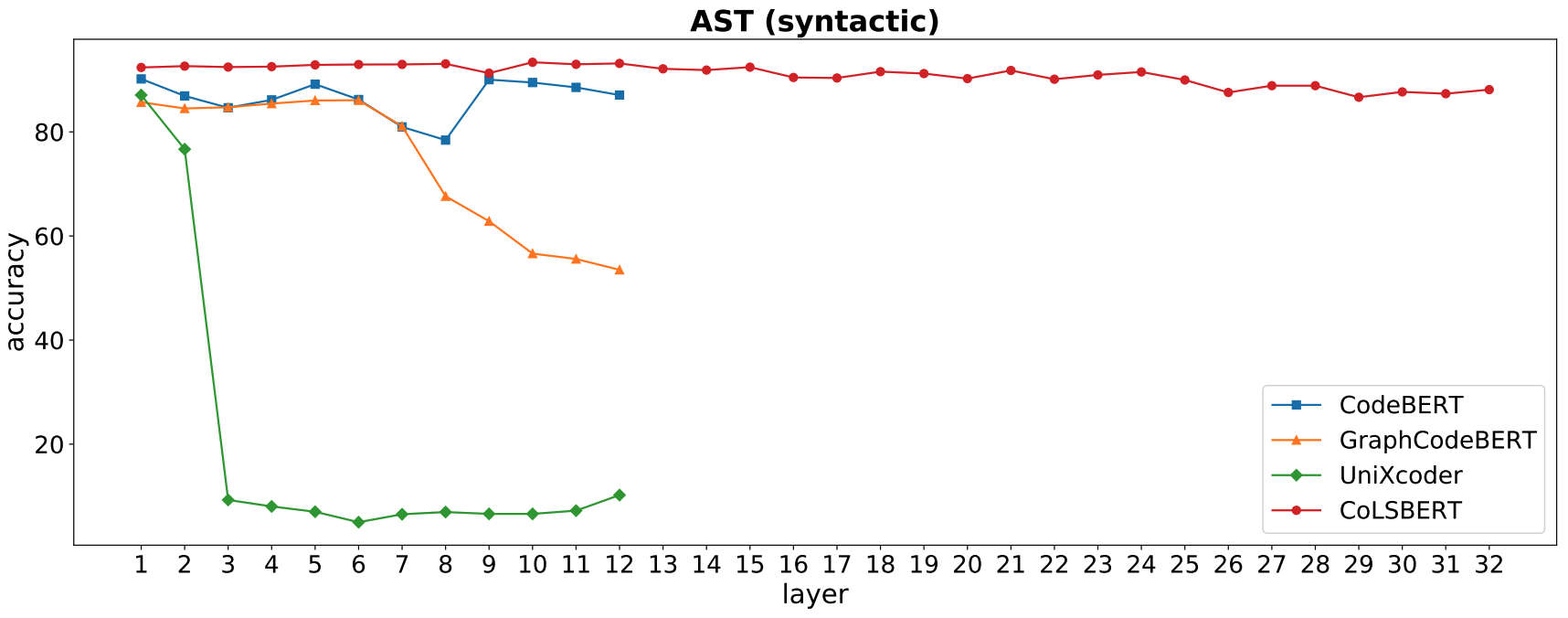}   
		\end{minipage}
	}
	{
		\begin{minipage}{0.48\textwidth}
			\centering      
			\includegraphics[scale=0.28]{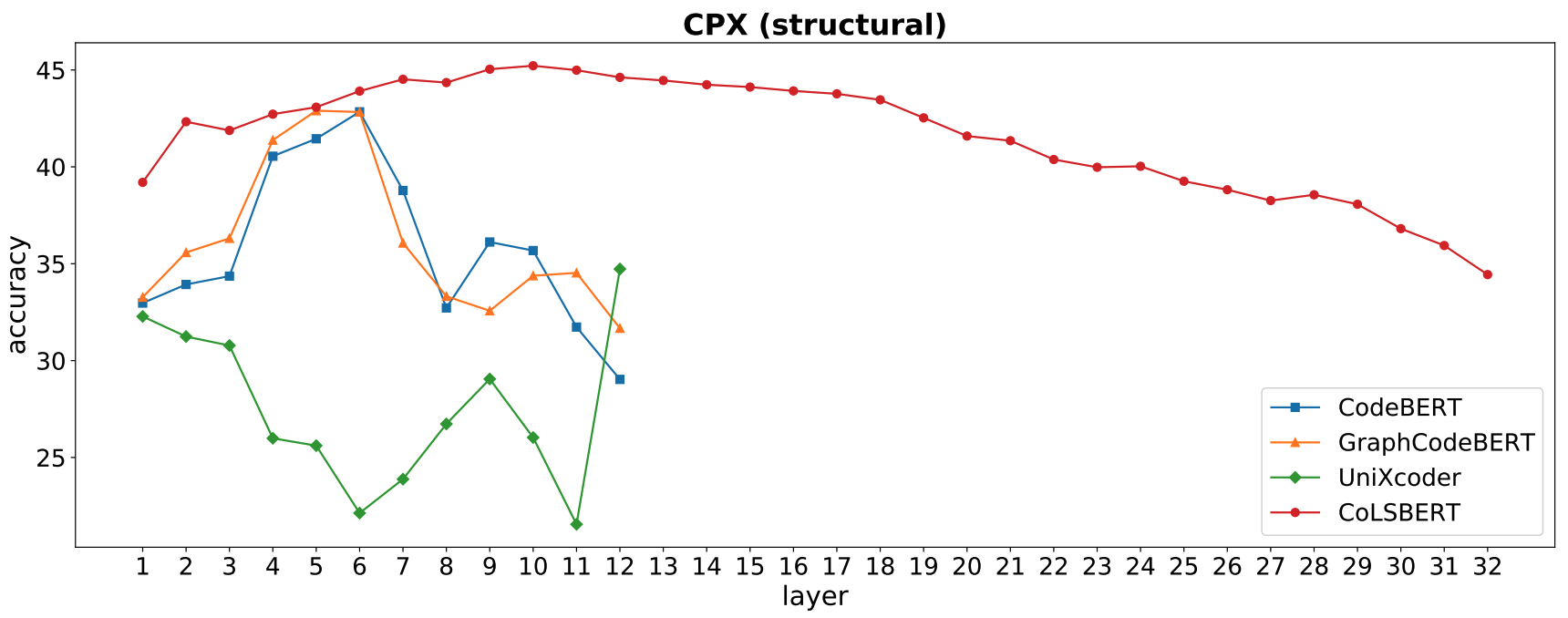}   
		\end{minipage}
	}
	{
		\begin{minipage}{0.5\textwidth}
			\centering      
			\includegraphics[scale=0.28]{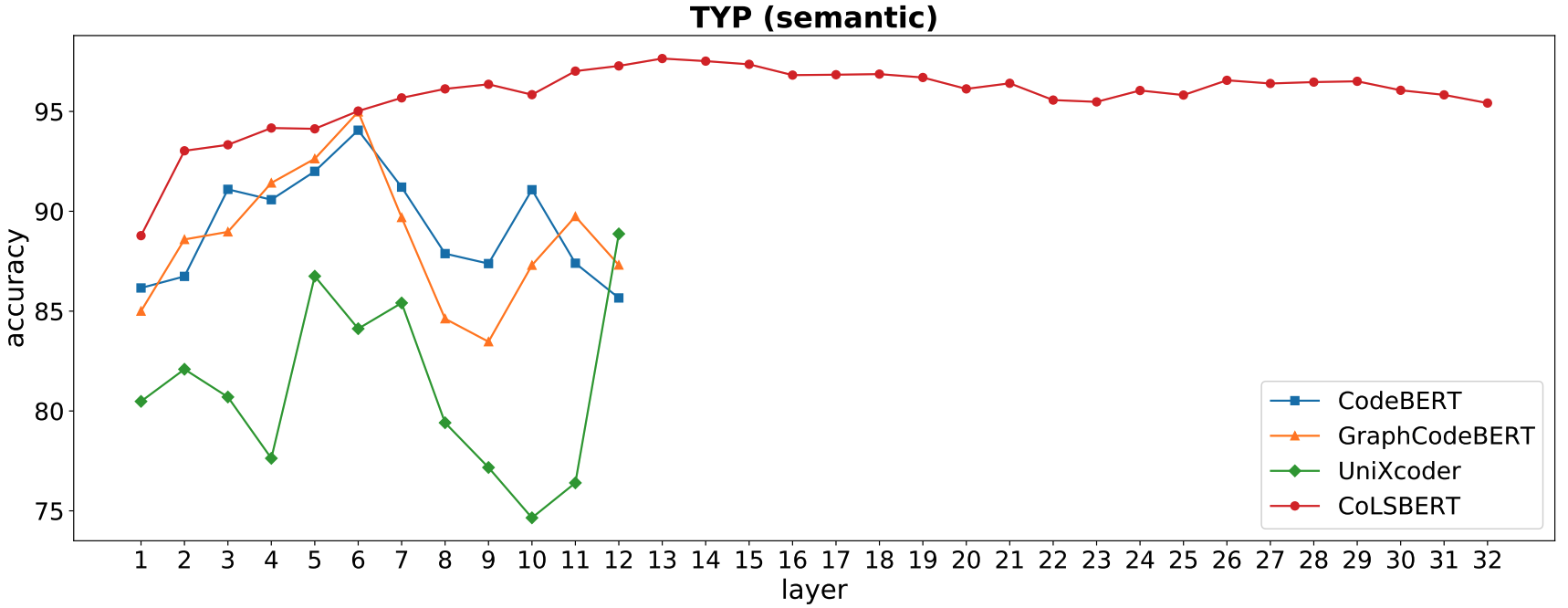}   
		\end{minipage}
	}
	\caption{Accuracy of different code search fine-tuned models on LEN, AST, CPX and TYP tasks. The horizontal axis indicates the index of the hidden layer used for probing.} 
	\label{probing every layer}  
    \vskip -0.2in
\end{figure*}

To delve more profoundly into the nuanced variations in the aptitude of the fine-tuned models for comprehending code, 
we conduct additional analyses by extracting feature vectors from each hidden layer for probing experiments. 
With the exception of CoLSBERT, which comprises 32 layers, the other models consist of 12 layers.

Figure~\ref{probing every layer} illustrates the accuracy of the fine-tuned code search model across all layers. 
Our observations reveal that all fine-tuned models exhibit heterogeneous performance patterns across layers, aligning with the findings of Karmakar and Robbes~\cite{DBLP:conf/kbse/KarmakarR21}. 
Notably, both CodeBERT and GraphCodeBERT showcase an initial increase followed by a subsequent decrease in accuracy across all tasks. 
Conversely, UniXcoder manifests a decline followed by an increase in accuracy for the LEN, CPX, and TYP tasks. 
However, concerning the AST task, a sharp accuracy drop occurs at the third layer, potentially indicative of adverse effects from other pre-training tasks.
In the case of CoLSBERT, accuracy diminishes layer by layer for the LEN task, while maintaining consistently high accuracy for the AST task. 
The CPX task's accuracy demonstrates an initial ascent followed by a decline, whereas the TYP task's accuracy exhibits a continuous upward trend across layers. 

This observation suggests that these models acquire surface and syntactic information at a shallow level, while delving into the structural and semantic aspects of code at deeper layers.

\subsection{Probing Results of the Code Clone Models}
\label{probing_code_clone}
\begin{figure*}[ht]	
    \vskip 0.2in
	{
		\begin{minipage}{0.48\textwidth}
			\centering          
			\includegraphics[scale=0.28]{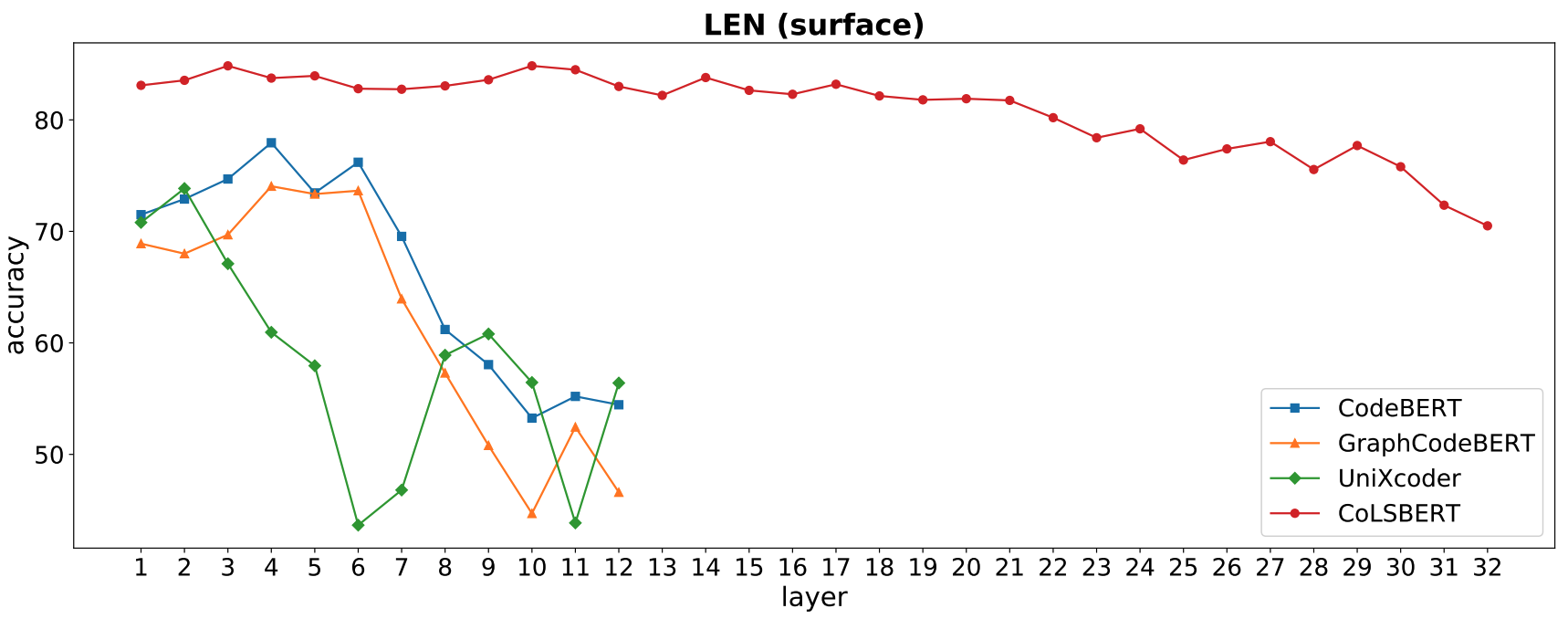}   
		\end{minipage}
	}
	{
		\begin{minipage}{0.5\textwidth}
			\centering     
			\includegraphics[scale=0.28]{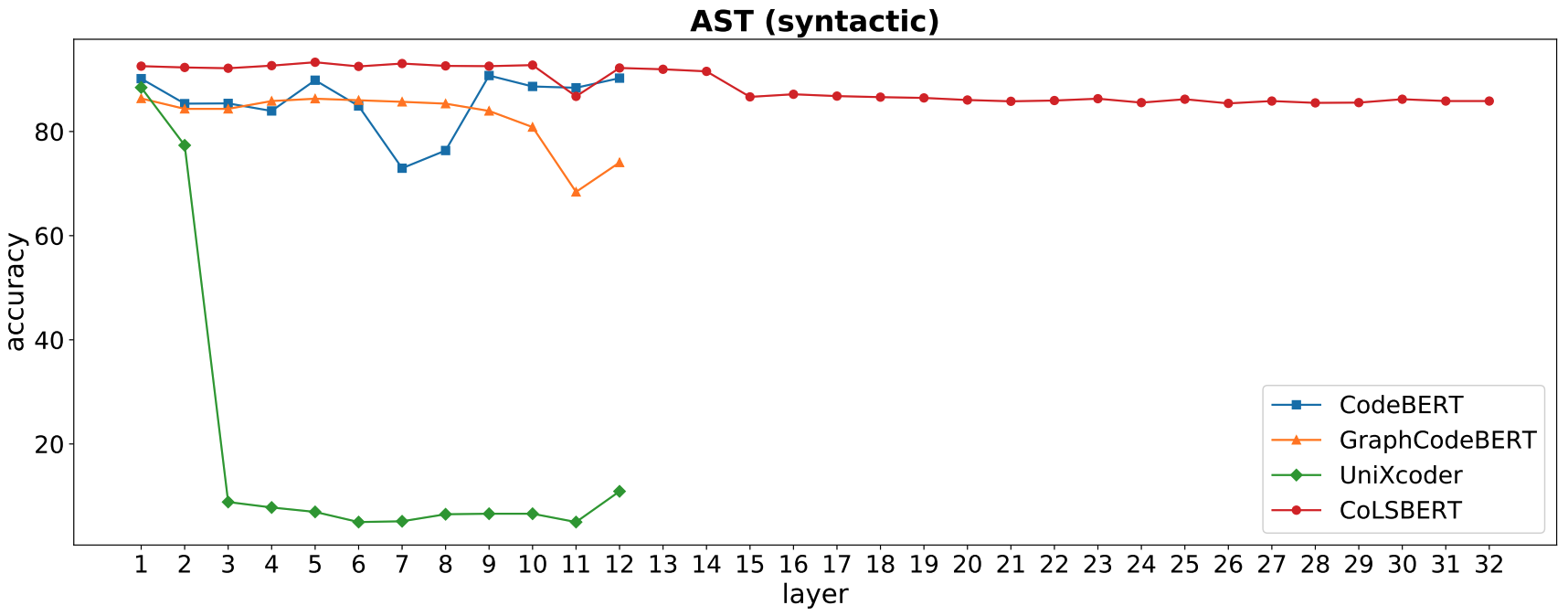}   
		\end{minipage}
	}
	{
		\begin{minipage}{0.48\textwidth}
			\centering      
			\includegraphics[scale=0.28]{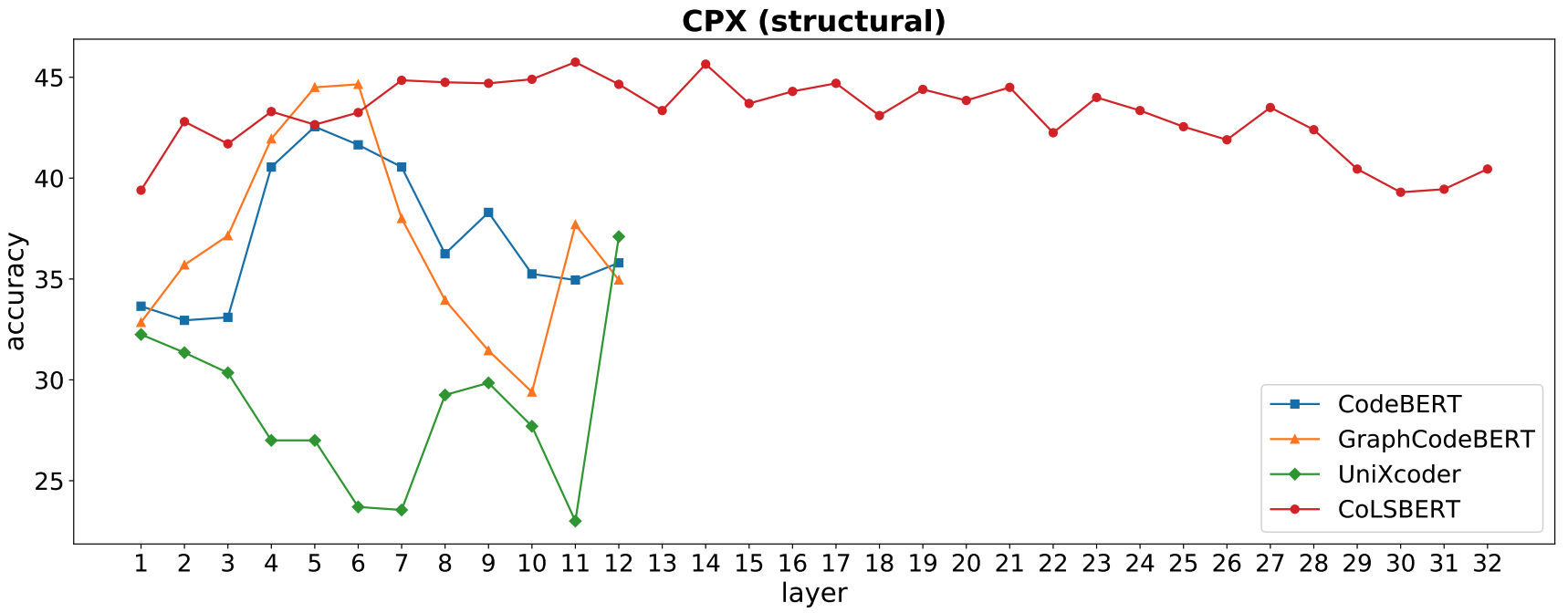}   
		\end{minipage}
	}
	{
		\begin{minipage}{0.5\textwidth}
			\centering      
			\includegraphics[scale=0.28]{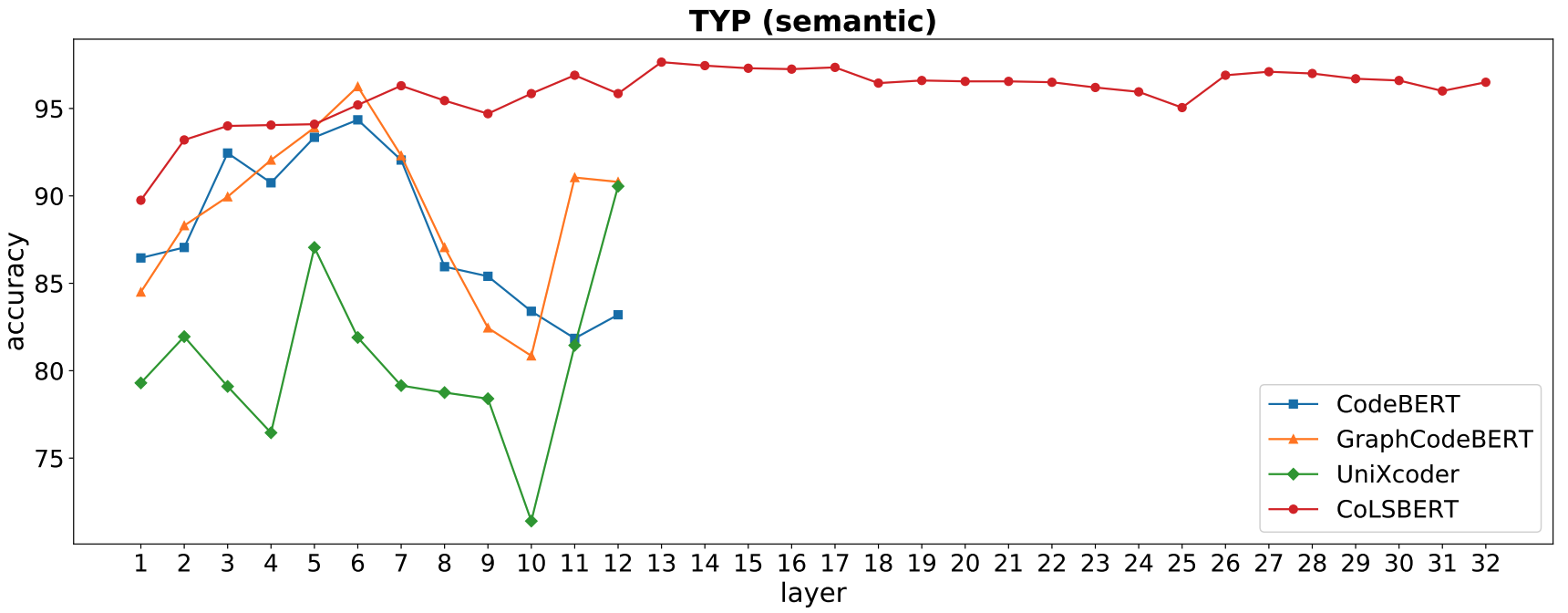}   
		\end{minipage}
	}
	\caption{Accuracy of different clone detection fine-tuned models on LEN, AST, CPX and TYP tasks. The horizontal axis indicates the index of the hidden layer used for probing.} 
	\label{CD probing every layer}  
    \vskip -0.2in
\end{figure*}

Table~\ref{CD_Probing} showcases the probing results of the clone detection models, demonstrating the superior performance of CoLSBERT in the LNE, CPX, and TYP tasks. 
This finding implies that CoLSBERT effectively captures surface, structural, and semantic information of the code within its hidden layers. 
Furthermore, across all four tasks, these clone detection models consistently outshine code search models. 
Figure~\ref{CD probing every layer} visually represents the probing results for feature vectors at each layer of these models, 
highlighting analogous trends in cross-layer accuracy changes across the four tasks, mirroring the patterns observed in the code search models.

\begin{table*}[ht]
\caption{Probing task accuracy on clone detection fine-tuned models.}
\label{CD_Probing}
\begin{center}
\begin{small}
\begin{sc}
\resizebox{0.7\textwidth}{!}{
\begin{tabular}{ccccc}
\toprule
\multirow{2}{*}{Model}                                &  LEN        &    AST       &     CPX         &  TYP       \\
\cmidrule(lr){2-5}
                                                        &  surface    &    syntactic &    structural   & semantic   \\
\midrule
CodeBERT~\cite{DBLP:conf/emnlp/FengGTDFGS0LJZ20}      &    54.45    &   \textbf{90.25} &    35.80      &    83.20       \\
\midrule
GraphCodeBERT~\cite{DBLP:conf/iclr/GuoRLFT0ZDSFTDC21} &   46.60     &    74.05       &     34.95       &    90.80       \\
\midrule
UniXcoder~\cite{DBLP:conf/acl/GuoLDW0022}             &   56.40     &   10.90      &     37.1       &    90.55         \\
\midrule
CoLSBERT                                              &   \textbf{70.50}     &    85.85     &     \textbf{40.45}        &    \textbf{96.5}          \\
\bottomrule
\end{tabular}
}
\end{sc}
\end{small}
\end{center}
\end{table*}

\end{document}